\documentclass[letterpaper,twocolumn,10pt]{article}
\usepackage{usenix-2020-09}

\usepackage{booktabs,xcolor,hyperref,enumerate}
\usepackage[T1]{fontenc}
\usepackage[utf8]{inputenc}
\usepackage{bm,courier,xspace,fancyhdr,fancyref,wrapfig}
\usepackage{amsmath,amsfonts,tabularx,tablefootnote,wrapfig,hyphenat}
\usepackage[labelformat=simple,skip=0pt]{subcaption}
\usepackage{xfrac,algorithm,algpseudocode}
\usepackage{paralist,mathtools,amsmath,upgreek}
\usepackage{float,enumitem,lipsum}
\usepackage[capitalize]{cleveref}
\usepackage{MnSymbol}
\usepackage{wasysym}
\microtypecontext{spacing=nonfrench}

\usepackage[toc]{appendix}

\newcommand{\parabreak}{\vspace*{2.00ex minus 0.25ex}\noindent}
\newcommand{\parahead}[1]{\vspace*{1ex plus 0ex minus 0.25ex}\noindent{}{\bfseries #1}}
\let\oldReturn\Return
\renewcommand{\Return}{\State\oldReturn}

\newcommand{\shortname}{mmWall}
\newcommand{\systemname}{\shortname{}}
\newcommand{\shortnames}{mmWall's}
\newcommand{\systemnames}{\shortnames{}}
\newcommand{\shortnamepl}{mmWalls}

\setitemize{itemsep=3pt,topsep=4pt,parsep=2pt,partopsep=0pt,leftmargin=1em}
\setenumerate{itemsep=3pt,topsep=4pt,parsep=1pt,partopsep=0pt,leftmargin=1em}

\crefformat{section}{\S#2#1#3}
\crefname{figure}{Fig.}{Figs.}
\crefformat{equation}{Eq.~(#2#1#3)}
\crefmultiformat{equation}{Eqs.~(#2#1#3)}{ and~(#2#1#3)}{, (#2#1#3)}{ and~(#2#1#3)}

\begin{document}

\thispagestyle{empty}

\title{mmWall: A Transflective Metamaterial Surface for mmWave Networks}
\author{{\rm Kun Woo Cho$^1$, Mohammad H. Mazaheri$^3$, Jeremy Gummeson$^2$, Omid Abari$^3$, Kyle Jamieson$^1$}\\ Princeton Univ.$^1$, Univ. of Massachusetts Amherst$^2$, UCLA$^3$}


\maketitle

\begin{abstract}
Mobile operators are poised to leverage 
millimeter wave technology as 5G evolves, but despite efforts 
to bolster their reliability indoors and outdoors, mmWave links
remain vulnerable to blockage by walls, people, and obstacles.
Further, there is significant interest in bringing outdoor mmWave 
coverage indoors, which for similar reasons remains challenging today.
This paper presents the design, hardware implementation, and 
experimental evaluation of \textit{\shortname{}}, the first
electronically almost-360$^\circ$ steerable
metamaterial surface that operates above 24~GHz and
both refracts or reflects incoming mmWave transmissions.  
Our metamaterial design consists of arrays of varactor\hyp{}split
ring resonator unit cells, miniaturized for mmWave.
Custom control circuitry drives each resonator, 
overcoming coupling 
challenges that arise at scale. Leveraging beam steering
algorithms, we integrate \shortname{} into the link layer discovery
protocols of common mmWave networks.  We have fabricated a 
10~cm by 20~cm \shortname{} prototype consisting of an
$28$ by $76$ unit cell array, and evaluate in
indoor, outdoor-to-indoor, and multi-beam scenarios.
Indoors, \shortname{} guarantees 91\% of locations
outage-free under 128-QAM mmWave data rates
and boosts SNR by up to 15~dB.
Outdoors, \shortname{} reduces the probability of complete link failure by 
a ratio of up to 40\% under 0--80\% path blockage 
and boosts SNR by up to 30~dB. 
\end{abstract}

\section{Introduction}
\label{s:intro}

Millimeter-wave
(mmWave) spectrum has emerged in the 5G/6G era as a key next generation wireless 
network enabler, fulfilling user demands for high spectral efficiency and 
low latency wireless networks. 
Higher carrier frequencies offer greater network capacity: for instance, 
the maximum carrier frequency of the 4G LTE band at 2.4~GHz provides an 
available spectrum bandwidth of only 100~MHz, while mmWave 
(above 24~GHz) can easily hold spectral bandwidths five to ten times greater,
enabling multi\hyp{}Gbit\fshyp{}sec data rates. 
Hence, mmWave spectrum
enables a plethora of mobile applications 
that are currently infeasible due to their requirements of very high
data rates, such as
virtual and augmented reality (VR\fshyp{}AR),
camera\hyp{}based purchase tracking in smart stores, 
and robotic automation in smart warehouses.

\begin{figure}
\centering{}\textbf{Without \shortname{}:}\qquad{}
\qquad{}\textbf{With \shortname{}:}\\
\centering
\begin{subfigure}[b]{1\linewidth}
\includegraphics[width=1\linewidth]{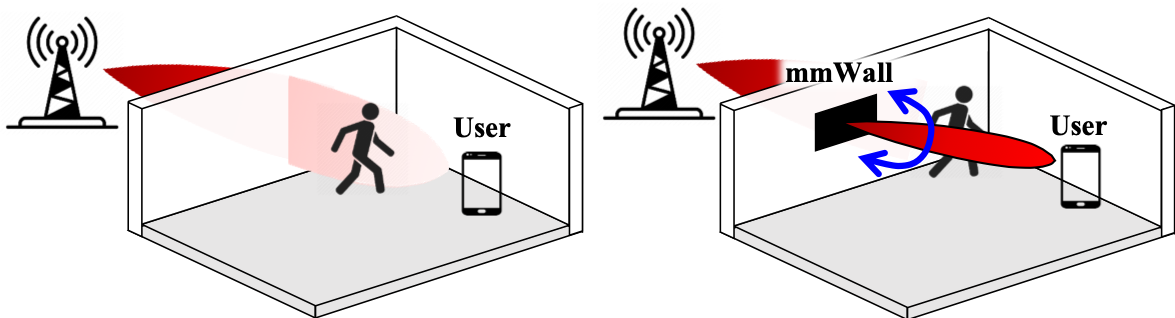}
\caption{5G/6G outdoor-to-indoor coverage via \shortname{}.}
\label{f:scenario1}
\end{subfigure}
\begin{subfigure}[b]{1\linewidth}
\hfill
\includegraphics[width=0.9\linewidth]{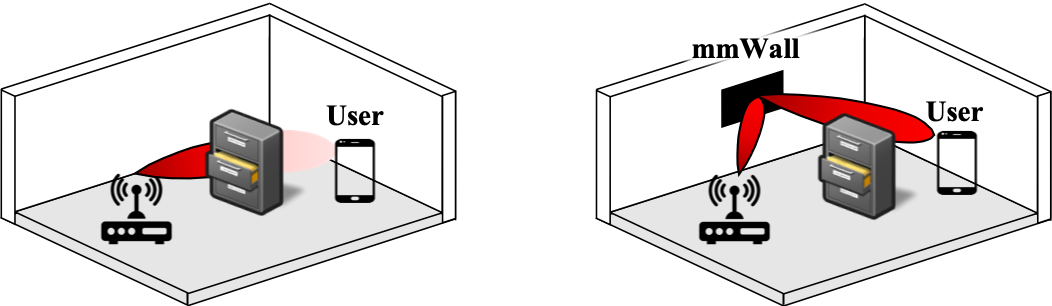}
\caption{\shortnames{} reflective mode for indoor VR/AR.}
\label{f:scenario2}
\end{subfigure}
\begin{subfigure}[b]{1\linewidth}
\includegraphics[width=1\linewidth]{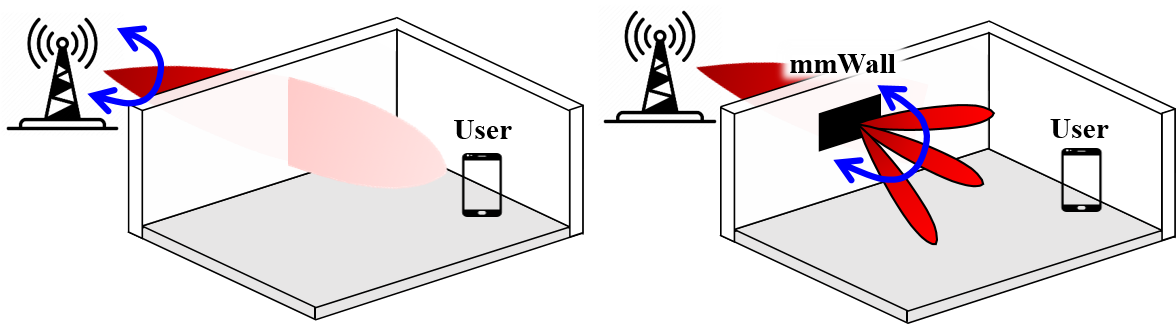}
\caption{\systemnames{} beam splitting, for link establishment.}
\label{f:scenario3}
\end{subfigure}
\caption{\shortname{} re-focuses outdoor coverage
indoors towards the user and potentially around obstacles,
provides path diversity indoors by reflection, and
splits an incoming beam for fast link establishment.}
\label{f:intro_use_cases}
\end{figure}

mmWave technology faces significant headwinds, however, in
at least three key scenarios:

\begin{enumerate}
\item Indoors, people, furniture, doors, and other clutter
block mmWave (\Cref{f:scenario1}),  
forcing data to flow over a much less reliable reflection path.
Indeed, in an extensive indoor measurement campaign at 28~GHz,
MacCartney \emph{et al.} observe a close-in
best non\hyp{}line of sight path loss exponent \emph{ca.} 3, 
with a normally\hyp{}distributed additional
loss with an 11~dB variance
\cite{7289335}.  While the resulting temporary outages 
are common, highly demanding
applications like VR/AR streaming cannot tolerate these glitches.

\item Second, 5G outdoor coverage is difficult to bring indoors,
as exterior building walls block
mmWave signal, as do outdoor windows' tinted glass (\Cref{f:scenario2}). 
Attenuation at 28~GHz is \emph{ca.} 40~dB 
versus 4~dB through
indoor glass \cite{6655403}, as outdoor
metalized glass coatings attenuate by 25--50~dB 
per layer \cite{3038}.
Currently, operators are forced to offload mmWave traffic 
onto lower frequencies or off their networks entirely 
(Wi-Fi) when users move indoors,  
incurring handover delay and application disruptions.

\item Third, NextG cellular providers face challenges in
adopting mmWave frequencies outdoors for primary service 
as well as wireless backhaul because
mmWave signals are readily absorbed by foliage, and 
reflection off buildings is largely specular, constraining 
the angle of reflection to be equal to the angle of incidence,
as shown in \Cref{f:scenario3}. 
Measurements in New York City highlight this issue:
28~GHz data shows most links greater than 
200~meters in outage \cite{6655399}.
\end{enumerate}

\parabreak{}This paper describes the design and implementation of
\textit{\textbf{\shortname{}}}, an electronically reconfigurable surface that addresses
all three foregoing use cases, also shown in Figure~\ref{f:intro_use_cases}.
Like much prior work (\S\ref{s:related}), \shortname{} leverages 
\emph{metamaterials}, artificial composite materials
engineered at a sub-wavelength scale to exhibit unique electromagnetic 
properties that do not exist in naturally occurring materials \cite{6230714}.
But \shortname{} is the first practical work to our knowledge 
to use a specific class of metamaterials capable of refracting 
incoming radiation with (theoretically) no loss: \emph{Huygens} metamaterials
\cite{6891256,PhysRevLett.110.197401}.
\shortname{} is a reconfigurable intelligent surface that 
uses a novel Huygens metasurface (HMS) metamaterial to
reflect\fshyp{}refract and precisely steer incoming mmWave beams 
towards desired directions, thus enhancing path diversity for
mmWave networks.  Work has shown
that surfaces that can steer incoming mmWave transmissions in this 
way have the potential to dramatically
improve spatial multiplexing \cite{8964330} and 
spectral efficiency \cite{9838886} of networks as a whole.  
Hence when obstacles like a human body or
outdoor foliage blocks the line of sight (LoS) or
non line of sight (NLoS) paths, \shortname{} can often 
provide an alternative path that is not a simple reflection
or a straight\hyp{}line transmission, and hence would not
otherwise exist.  In the first scenario above, \shortname{}
reflects mmWave beams at non\hyp{}specular angles (those 
for which the angle of reflection is not equal to the 
angle of incidence).  In the second scenario, \shortname{}
can refract mmWave signals from outdoors to steer them
directly towards an indoor receiver, making outdoor
to indoor communication possible.  And in the third scenario,
\shortname{} can reflect outdoor transmissions at 
non\hyp{}specular angles, ameliorating outdoor 
blockages.

\shortname{} is electronically reconfigurable 
to either reflect or refract incoming energy, allowing it
to time\hyp{}multiplex the different roles of each of the three above
use cases, while installed in a fixed location.  
Also, its multi-beam functionality enables 
fast beam search, and support for multiple users at the same time.
To our knowledge, \shortname{} is the first surface 
able to achieve near\hyp{}$360^{\circ}$ angular coverage (\Cref{s:eval}).

\begin{figure*}
\centering
\includegraphics[width=1\linewidth]{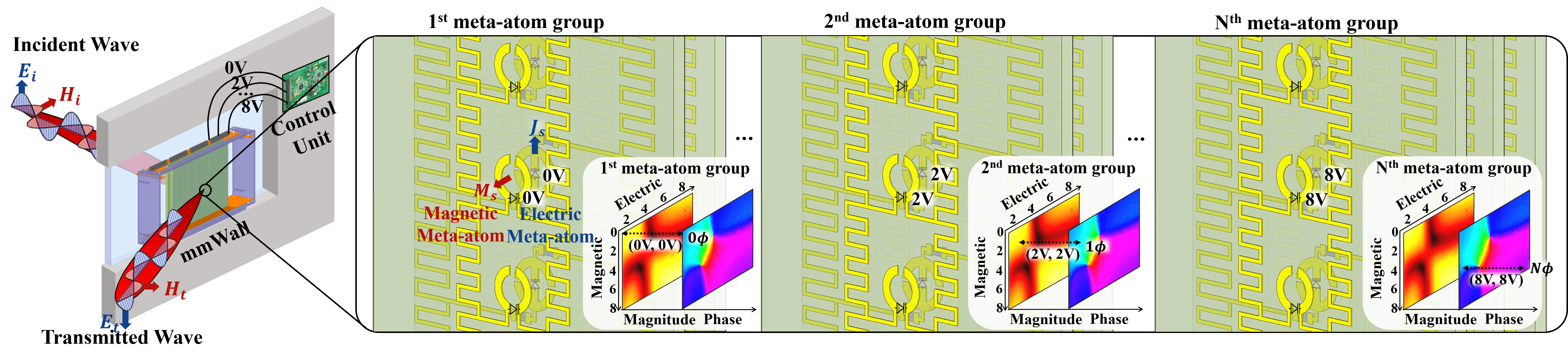}
\caption{\shortnames{} design converts an incident mmWave beam
to a refracted (or reflected, not shown) beam via
field discontinuities created by current
in its resonators. \textit{Inset:} electric meta\hyp{}atoms are shown
in front of the magnetic meta-atoms.}
\label{f:high_level_design}
\end{figure*}

This work addresses several hardware and software design 
challenges that arise in the realization of such a design.
Since mmWave transmissions are ``pencil-beam'' in nature, 
they work only when the transmitter’s beam 
is perfectly aligned with the receiver’s beam.
To correctly steer the beam towards the receiver, 
we design a metamaterials-based surface 
that can precisely control the phases of the incoming signal,
focusing signal power in a narrow beam.
Secondly, since the size of meta-atom scales with its operating frequency, 
\shortname{}'s meta-atoms are much smaller than the conventional antennas 
and therefore extremely sensitive to coupling.
Hence, we not only scale the surface to mmWave frequency 
but also deliberately design the control lines to avoid undesirable coupling.
Lastly, existing systems uses their own beam searching protocol to 
find the best alignment.
To make \shortname{} compatible with different mmWave applications,
we design an effective beam alignment protocol 
that leaves the existing systems unchanged \cite{hassanieh2018fast}.

\parahead{Contributions and Results.}
We analyze our meta-atom designs and compare 
them with simulation results, allowing our designs to scale to different 
frequencies for potential applications like Terahertz communication. 
To the best of our knowledge, this is the first study that 
theoretically analyzes and builds a working prototype of
a reconfigurable Huygens metasurface at mmWave frequency.
We have designed and implemented \shortname{} hardware in custom PCB,
and in \cref{s:eval}, evaluate its performance through experiments 
in environments matching the three scenarios we outline above.
Our empirical results show that when both the AP and the client 
are in the same room, we can provide an SNR of 25~dB or more for all 
locations in a $10 \times 8 m$ room, using a single \shortname{} surface.
This SNR is sufficient to support 128\hyp{}QAM in 91\% of locations. Moreover, we 
show that the SNR improves to 30 and 35~dB when we place two
surfaces, respectively, on different walls. Finally, we show the 
effectiveness of \shortname{} in bringing outdoor mmWave networks 
indoors.  In particular, when the AP is 6~meters away from the 
building, \shortname{} improves the SNR by up to 30~dB, 
providing an SNR of 20~dB or more in all locations in a room using 
a single surface placed on a window.

\section{Related Work}
\label{s:related}

Prior work in passive 
HMSs \cite{6891256, smith2004metamaterials, ding2020metasurface} 
has demonstrated a ``lensing'' effect and negative refraction index 
\cite{smith2004metamaterials} and the engineering of complex
beam patterns \cite{ding2020metasurface}.  
Prior work in actively\hyp{}controlled HMSs \cite{chen2017reconfigurable, 
zhang2018space, liu2018huygens, wu2019tunable, 9020088} 
uses varactors or PIN diodes to tune each element 
in a continuous or binary (i\emph{i.e.}, on\hyp{}off) manner, respectively.
Such active HMSs can shift incoming signals' frequency
\cite{liu2018huygens} and polarization~\cite{chen2020pushing, wu2019tunable}.
While these designs have shown great promise in theoretical 
prediction models \cite{9514544}
and\fshyp{}or at frequencies
below \emph{ca.} 10~GHz, they 
do not scale to higher mmWave frequencies in a straightforward way,
due to a mismatch between the 
required meta\hyp{}atom size and a varactor's size, and the attenuation
that commonly available substrates would induce on an incident mmWave
signal, and so do not address the use cases \shortname{} targets.
\shortname{} is the first mmWave work to do so.
Evaluation efforts 
in this group of prior work stop short of realistic
end\hyp{}to\hyp{}end experiments.

Basar \emph{et al.} \cite{8796365} and Liu \emph{et al.} \cite{9424177} 
survey the area of Reconfigurable Intelligent Surfaces (RISs) 
in general (concentrating on lower,
sub\hyp{}6~GHz bands), the latter discussing their interactions with
ML algorithms, unmanned aerial vehicles, and other technologies 
in the 6G roadmap.  Representative work in the 2.4~GHz Wi-Fi
context has studied the transmissive, through\hyp{}wall
scenario \cite{li2019towards} as well as the reflective
scenario \cite{arun2020rfocus, DBLP:conf/mobicom/DunnaZSB20}.
Path modelling efforts apply
well\hyp{}established radio propagation principles on the
radio propagation channel to calculate link budgets 
for RIS hardware \cite{9206044}.

Work in actively\hyp{}controlled mmWave RISs includes 
a solely reflective, PIN\hyp{}diode based
surface at 2.3 and 28~GHz \cite{9020088}, whose evaluation at 28~GHz
states a gain of 19~dBi, but which stops short of further experimental 
evaluation of steerability or any further 
end\hyp{}to\hyp{}end evaluation at 28~GHz.
Tang \emph{et al.} describe similar PIN\hyp{}diode, reflective
surfaces at 27 and 33~GHz, model path losses in such scenarios, and 
experimentally evaluate \cite{9837936}.  Tan \emph{et al.}
consider a similar design at 60~GHz \cite{tan2018enabling}, 
but neither consider 
HMS\hyp{}based designs such as \shortnames{}, which can 
shift between reflective (on both sides of the surface) 
and transmissive modes instantly via electronic control.
In press releases 
(\href{https://www.docomo.ne.jp/english/info/media_center/pr/2019/0529_00.html}{[a]}, \href{https://www.docomo.ne.jp/english/info/media_center/pr/2021/0126_00.html}{[b]},
\href{https://www.businesswire.com/news/home/20181204005253/en/NTT-DOCOMO-and-Metawave-Announce-Successful-Demonstration-of-28GHz-Band-5G-Using-Worlds-First-Meta-Structure-Technology}{[c]}) 
NTT DoCoMo describe reflective, outdoor\hyp{}to\hyp{}indoor
surfaces  operating at 28~GHz.  They state top line experimental
results, but do not disclose design details 
or details of their experimental evaluation.
Other work uses split ring resonators as antennas
for a Massive MIMO base station \cite{9324910}, a related
but distinct application to \shortname{}.  This paper
is an extension of the authors' previous workshop publication
\cite{anonymous} that describes a new control line design,
documents real hardware implementation, and presents significant
new evaluation results in realistic, diverse scenarios.

Recent work in passive non\hyp{}Huygens metamaterials based mmWave RISs
includes proposals that reflect mmWave signals at angles of reflection
different than incidence \cite{millimirror-mobicom22, esmail2019new}, 
but cannot be tuned to target a receiver's location, hence wasting 
incident energy and resulting in at
most 10~dB of gain, significantly below \shortnames{} achieved gain.  
Also these approaches do not refract as \shortname{} can, 
yielding reduced applicability relative to \shortname{}.

Recent amplify-and-forward proposals for Wi-Fi \cite{10.1145/3452296.3472890}
use a mesh topology, but 
do not scale to mmWave frequencies, and at mmWave \cite{abari2017enabling} 
are limited to indoor reflection.
Recent complementary approaches 
leverage multi-beam transmission 
\cite{jog-nsdi19, jain-sigcomm21},
sensing and leveraging ambient reflectors
\cite{wei-nsdi17}, and use Wi-Fi as a control plane to 
discover mmWave links
\cite{sur-mobicom17, 9709079}.  While they align
with \shortnames{} goals, such approaches cannot create
paths whose reflection angles diverge from
their incident angles, or refract through a surface.
\begin{figure*}[t]
\begin{subfigure}[b]{.33\linewidth}
\centering
\includegraphics[width=1\linewidth]{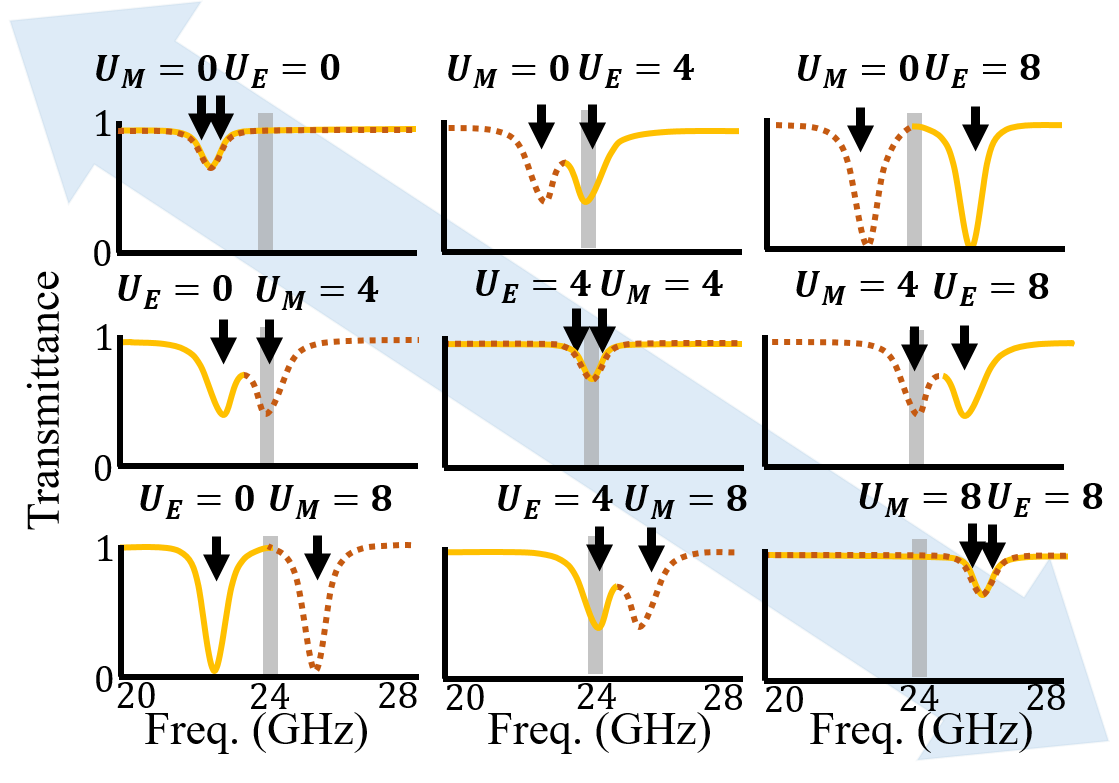}
\caption{Transmission magnitude $|T|$}
\label{f:huygen_mag}
\end{subfigure}
\begin{subfigure}[b]{.33\linewidth}
\centering
\includegraphics[width=1\linewidth]{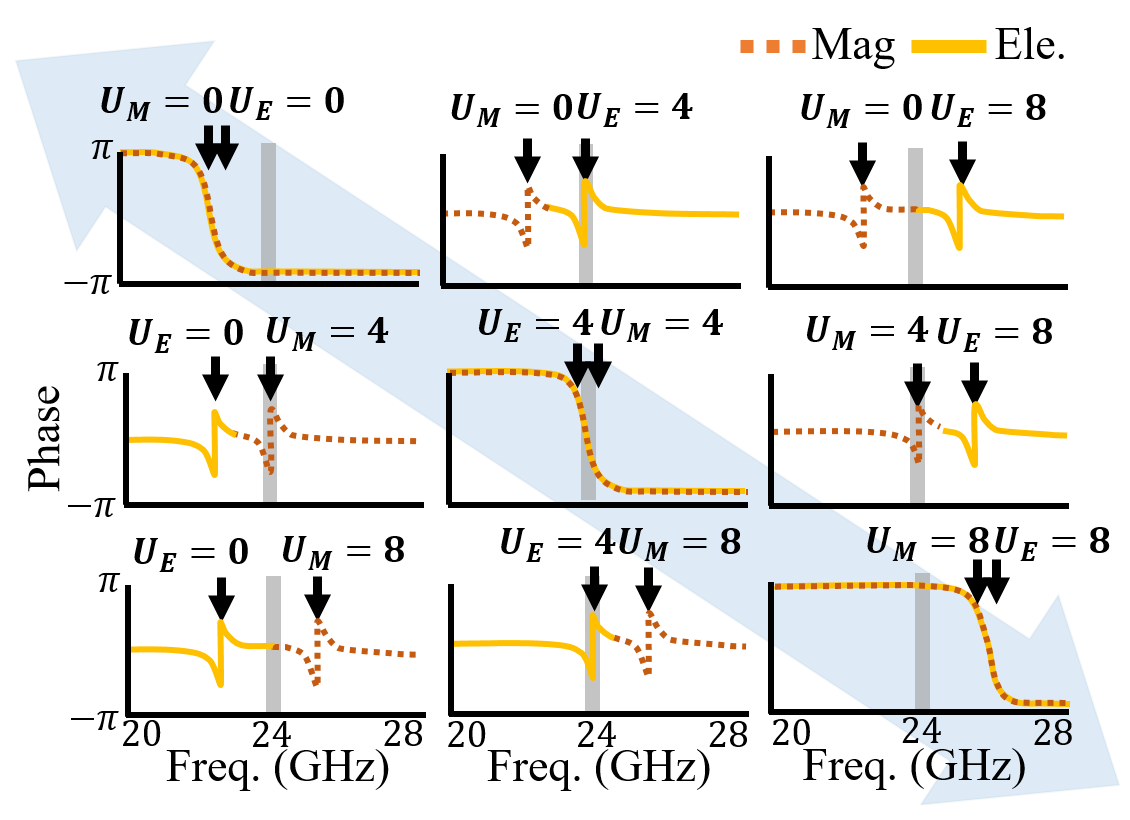}
\caption{Transmission phase shift $\angle T$}
\label{f:huygen_phase}
\end{subfigure}
\begin{subfigure}[b]{.33\linewidth}
\centering
\includegraphics[width=1\linewidth]{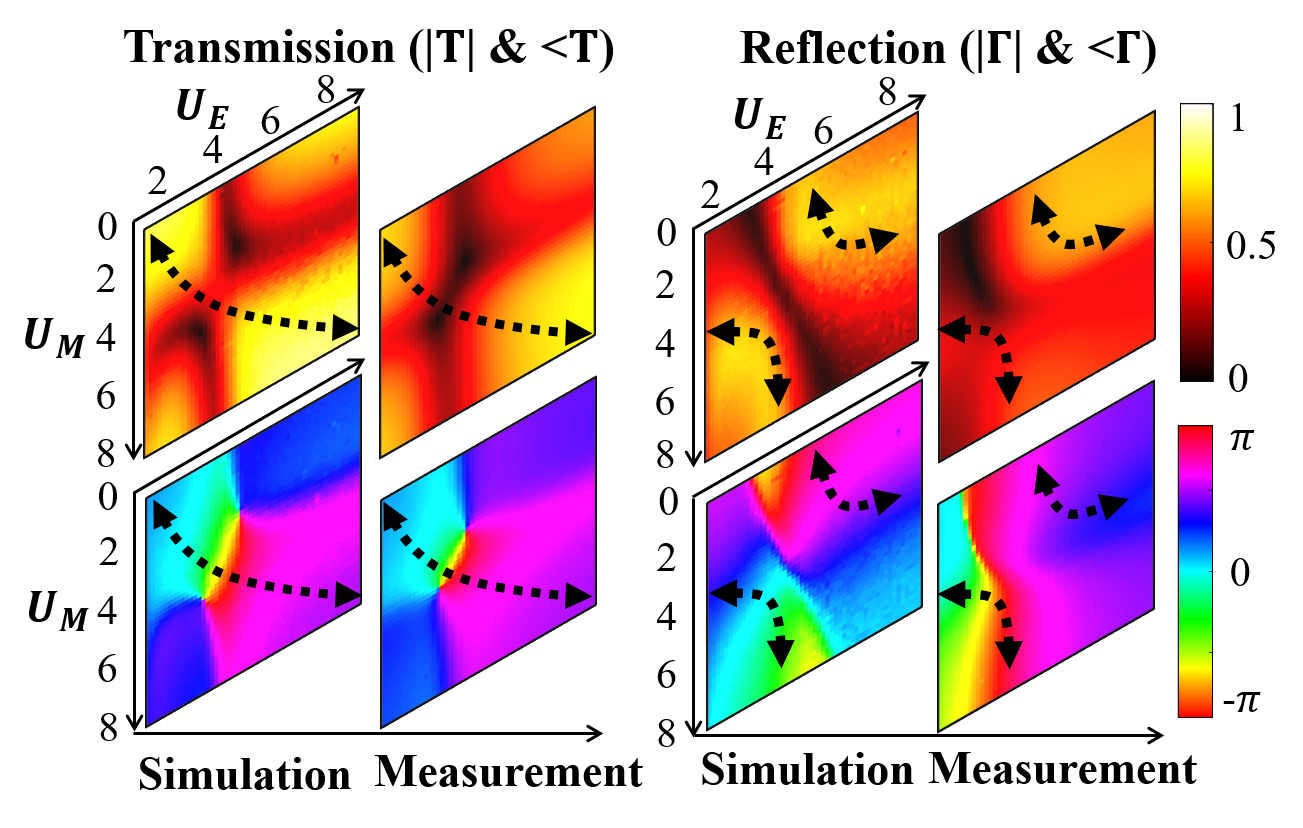}
\caption{$T$ and $\Gamma$ magnitude, phase at 24~GHz.}
\label{f:huygen_heatmap}
\end{subfigure}
\caption{Unit cell response \emph{v.} electric\hyp{} and
magnetic\hyp{}side control voltages $U_{E}$ and 
$U_{M}$---\textbf{(a):}~magnitude and \textbf{(b):}~phase.
\textbf{(c):}~HFSS simulation (\emph{left}) and 
near\hyp{}field, real world VNA measurement (\textit{right})--- arrows 
indicate control voltage pairs that yield a 360$^{\circ}$ phase
shift of the incoming signal, with high transmission 
or reflection magnitude.}
\label{f:huygen_pattern}
\end{figure*}

\section{Design}
\label{s:design}

We describe in turn \shortnames{} unit cells (\S\ref{s:design:unit}),
their control mechanism (\Cref{s:design:control}), and
their link layer integration (\Cref{s:design:link}).

\subsection{Unit Cell}
\label{s:design:unit}

\systemnames{} unit cells (also known as \emph{meta-atoms}) are stacked
vertically with a $\lambda/3$ separation, on each Rogers substrate
board (also known as a \emph{meta-atom group}), as shown
in \Cref{f:high_level_design}
(see \Cref{s:design:control:line} for a discussion of vertical and
horizontal unit cell spacing considerations for beamforming).

\subsubsection{Design Goals}
\label{s:design:unit:goals}

The two primary design goals of the unit 
cell are to simultaneously \textbf{1)}~achieve transmission $T$ or
reflection $\Gamma$ loss levels as close to zero as possible,
and \textbf{2)}~effect any phase shift in $[0,2\pi]$ on the
incoming signal, both at mmWave frequencies.
The unit cell consists of two meta-atoms, \emph{magnetic} and 
\emph{electric}, etched onto the two respective sides of a Rogers substrate
(\Cref{f:high_level_design}, \emph{inset}).
The magnetic (electric) meta-atom induces a magnetic (electric)
field response to the incoming
signal that can resonate at different, tunable
frequencies by varying the applied voltage to the varactor of
the magnetic\hyp{} (electric\hyp{}) side 
meta\hyp{}atom (see \Cref{f:high_level_design}, and 
\Cref{s:design:control} next). 

Without loss of generality, we now describe how transmission works
(reflection is fully complementary to transmission, and we
refer the reader to \Cref{s:theoretical_analysis} for a rigorous mathematical 
exposition of both).  
In \Cref{f:huygen_mag}, we observe that increasing the voltage applied to 
the magnetic meta\hyp{}atom $U_{M}$ from 0~to 8~V (down the
three leftmost subplots) 
shifts its resonance frequency (lowest transmission magnitude
point of the red dotted line\footnote{In operation we largely avoid 
the lowest transmission nulls.}) to the right (we will analyze 
how this frequency shifting works in \Cref{s:design:unit:params}). 
Similarly, the electric meta-atom induces an electric response and its 
resonant frequency can be shifted by its own varactor
(reading similarly across the three topmost subplots).
Together their effects are superposed and we manipulate the collective 
magneto\hyp{}electric response that interferes with the incident 
plane wave.

The key characteristic that allows near\hyp{}perfect amplitude with 
full phase coverage appears when the two responses overlap at the same frequency.
Otherwise, the phase response undergoes a sharp change of only $\pi$ and its 
magnitude dips to nearly zero at its resonant frequency, as we see 
in \Cref{f:huygen_mag} and \Cref{f:huygen_phase} when the voltages 
applied to the magnetic and electric meta-atom differ by 8~V. 
However, as the two resonances start to overlap, transmission loss
decreases and the 
phase shift becomes $2\pi$ (on\hyp{}diagonal 
sub\hyp{}figures, \Cref{f:huygen_mag} and \Cref{f:huygen_phase}).
As a result, we achieve $2\pi$ phase coverage with near\hyp{}unity 
magnitude by increasing the voltage 
applied to both the magnetic and electric meta\hyp{}atoms together 
(\Cref{f:huygen_heatmap}, at control voltages 
indicated by the black curves). 

While the overlapped resonances can reach a perfect unitary
transmission magnitude in theory,
the Huygens pattern from our measurement shows a lower transmission 
magnitude on the area where abrupt phase shifts occur due to various 
reasons, including the sensitivity at mmWave frequency, fabrication loss, 
and measurement errors.






\subsubsection{Design Process}
\label{s:design:unit:params}

\begin{figure}[t]
\begin{subfigure}[b]{.47\linewidth}
\centering
\includegraphics[width=1\linewidth]{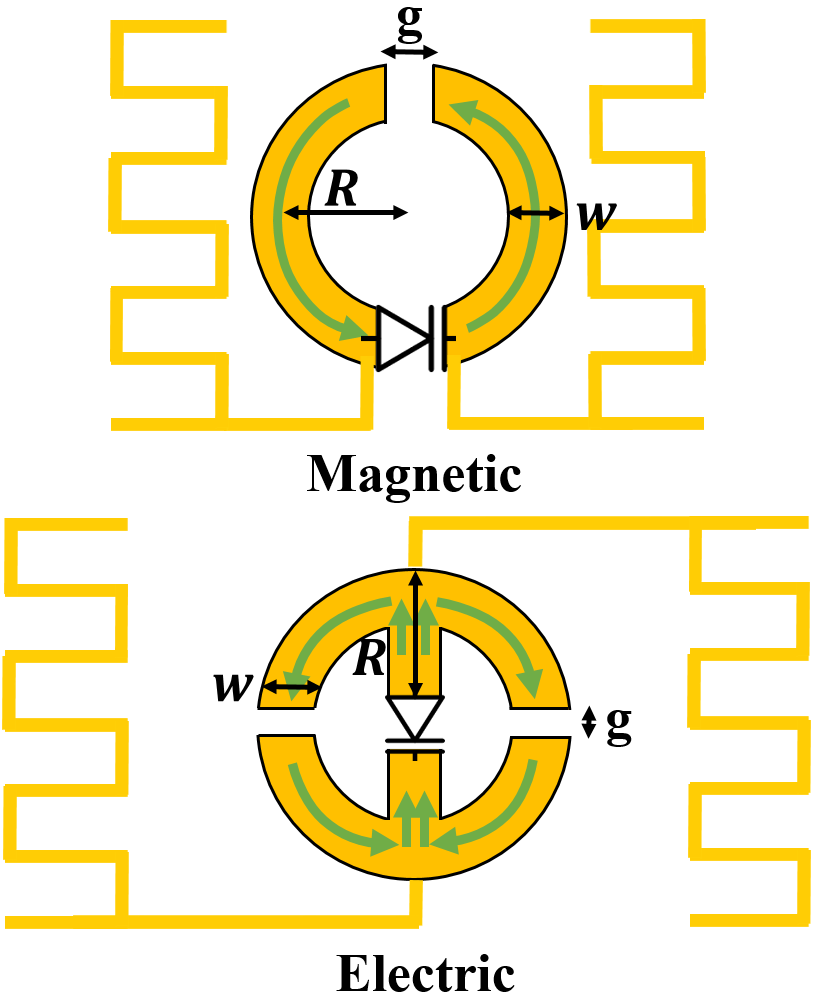}
\caption{\systemname{}}
\label{f:mmwall_unit_cell_design}
\end{subfigure}
\begin{subfigure}[b]{.26\linewidth}
\centering
\includegraphics[width=1\linewidth]{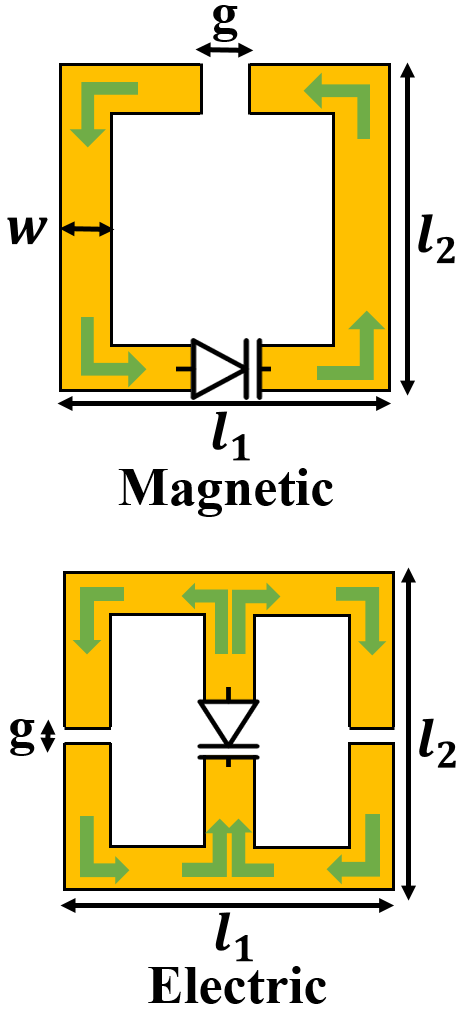}
\caption{Huygens}
\label{f:huygens_unit_cell_design}
\end{subfigure}
\begin{subfigure}[b]{.25\linewidth}
\centering
\includegraphics[width=.88\linewidth]{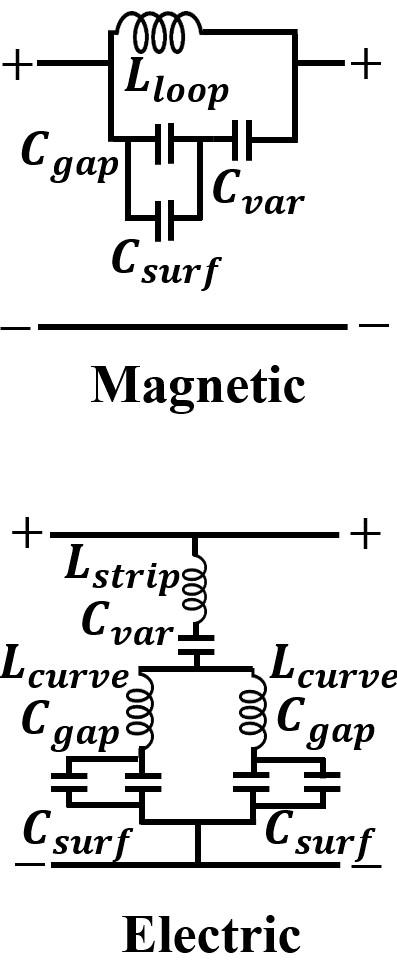}
\caption{Schematic}
\label{f:circuit_diagram}
\end{subfigure}
\caption{\shortname{}, prior Huygens unit cell 
designs (\emph{top:}~magnetic; \emph{bottom:} 
electric side), and equivalent circuits.}
\label{f:designs}
\end{figure}

We now describe challenges we overcame in scaling the resonance of the
\shortname{} unit cell to mmWave frequencies.
By definition, the meta-atom behaves as an $LC$ circuit with 
resonant frequency $1/\left(2\pi\sqrt{LC}\right)$, 
determined by the capacitance or inductance of the meta\hyp{}atoms.
Hence, we must markedly \emph{decrease} the inductance and capacitance
of prior microwave designs (\Cref{s:related}), 
if we can hope to achieve a mmWave resonant frequency.
As we will see next, the smaller the ring is, the higher 
the resonant frequency becomes.
However, the state-of-the-art approach to scale the frequency of a
Huygens resonator (\Cref{f:huygens_unit_cell_design})
requires a loop width $l_{1}$ and loop height $l_{2}$ of $\lambda/10$.
At mmWave, however, the varactor packaging itself
would significantly distort the tailored electromagnetic surface properties
when a meta-atom is sized $\lambda/10$, and so the straightforward
approach fails.

We thus instead adopt the design
shown in \Cref{f:mmwall_unit_cell_design}, but 
this is only tenable with a careful tradeoff of meta-atom
design parameters \textit{radius} $R$, \textit{trace width} $w$, and 
\textit{trace gap width} $g$ (\textit{cf.}~\Cref{f:designs}) as
we next describe.

\parahead{Magnetic meta-atom.} \Cref{f:mmwall_unit_cell_design} (\textit{upper})
shows the design parameters that determine inductance $L_m$ and capacitance $C_m$.
$L_m$ ($=L_{loop}$, the inductance of the physical conductor loop), 
is largely proportional to $R$ (also $L_{\mathrm{loop}} \propto$ 
$t^{-1}$, $w^{-1}$, and $g^{-1}$).
$C_{m}$ consists of three capacitance values, 
$C_{\mathrm{gap}}$, $C_{\mathrm{surf}}$, and $C_{\mathrm{var}}$:
\begin{equation}
    C_{m} = \left(\frac{1}{C_{\mathrm{gap}} + C_{\mathrm{surf}}} + 
        \frac{1}{C_{var}}\right)^{-1}
    \label{eq:magcap}
\end{equation}
Here, $C_{\mathrm{gap}}$ is the parallel\hyp{}plate capacitance induced by 
the gap in the ring ($\propto g^{-1}$),
$C_{\mathrm{surf}}$ is a capacitance induced by the metallic surface 
($\propto R$ \cite{vallecchi2019analytical}), and
$C_{\mathrm{var}}$ is the capacitance of the varactor, a voltage\hyp{}dependent 
capacitor.  While $L_{\mathrm{loop}}$, $C_{\mathrm{gap}}$, and $C_{\mathrm{surf}}$ 
are fixed after fabrication, $C_{\mathrm{var}}$ varies with control voltage.
Increasing $U_M$ decreases $C_{\mathrm{var}}$ (see \Cref{f:cvar_v_control} in
\Cref{s:unit_cell_analysis} for the precise relationship), and thus $C_{m}$ 
(\Cref{eq:magcap}), which in turn increases
the resonance frequency, as depicted in \Cref{f:huygen_pattern}.  

\begin{figure}
\centering
\includegraphics[width=1\linewidth]{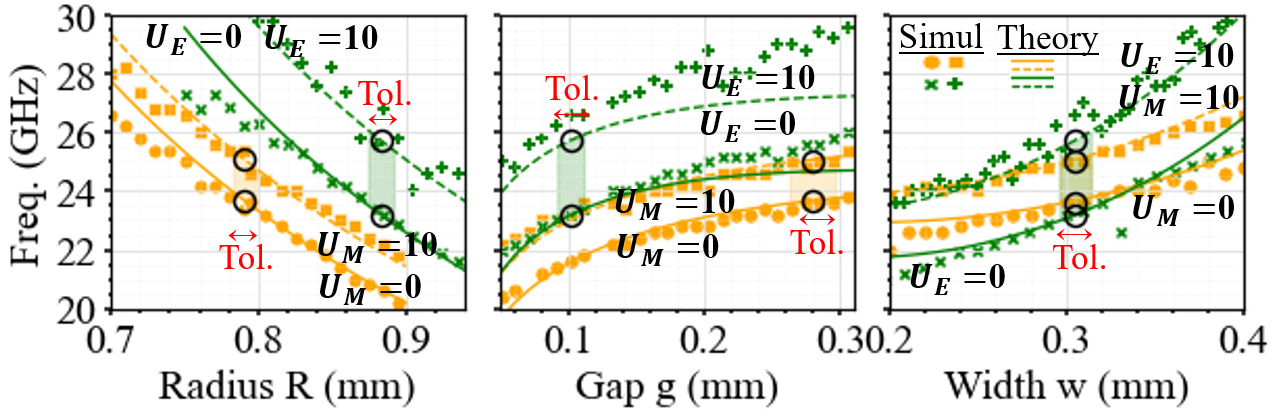}
\caption{\shortname{} design parameter sensitivity analysis.}
\label{f:theory:freq}
\end{figure}

When tuning the physical loop design parameters, we 
fix $C_{\mathrm{var}} = 4$~V for both the
magnetic and electric meta-atoms since at that voltage, the resonant frequency 
is at our desired mmWave frequency and an abrupt phase change occurs. 
\Cref{f:theory:freq} (curves labeled $U_M$) shows
our chosen design parameters (denoted with 
black circles) and its corresponding magnetic side 
resonant frequency when $U_M = 0, 10$~V.
Data in our sensitivity analysis (\Cref{f:theory:freq}) show 
that among all feature dimensions, decreasing $R$, followed by 
increasing $g$ has the greatest effect on increasing resonant 
frequency for the magnetic meta-atom.
We note that after fixing our meta-atom geometry as shown in the figure, 
24~GHz lies in the middle of the resulting resonant frequency range.
Also, we observe that PCB manufacturing tolerance ($\pm5\%$) does not greatly 
shift the resonant frequency
(we refer \Cref{s:vna} for meta-atom sensitivity analysis against fabrication 
tolerance).

\parahead{Electric meta-atom.}  \Cref{f:designs} (\emph{lower}) shows
the electric meta\hyp{}atom, in which
current oscillates in two different directions,
while the current of the magnetic meta-atom oscillates in one direction only
(\emph{cf.} green arrows in 
\Cref{f:mmwall_unit_cell_design,f:huygens_unit_cell_design}).
Hence, we analyze its inductance $L_e$ as the combination of the inductances
of the half\hyp{}circular loop on the left half ($L_{curve}$), the inductance of the 
other half on the right half ($L_{curve}$, by symmetry), 
and the inductance from the metallic strip shared by two loops ($L_{strip}$).
Since the two half\hyp{}loops are arranged in parallel, with 
the metallic strip  arranged in series,
$L_{e} = (L_{curve}/2) + L_{strip}$ \cite{esmail2019new}.
Since inductance generally depends on the surface area of the copper trace,
$L_{curve} \propto R$, and $L_{strip} \propto w^{-1}$, 
$L_{e}$ largely depends on both $R$ and $w$, but not $g$.
We see the impact of $w$ on the resonant frequency in 
\Cref{f:theory:freq}: compared to magnetic meta\hyp{}atom, the resonant 
frequency of the electric meta-atom increases steeply as $w$ increases due 
to $L_{strip}$.  To minimize the difference in resonant
frequencies between the electric and magnetic sides as desired,
\Cref{f:theory:freq} guides us to design an electric meta-atom with equal 
$w$ as the magnetic meta-atom, greater $R$ and lesser $g$. 
The electric meta-atom has two gaps and two surface capacitances,
with respective associated capacitances 
$C_{\mathrm{gap}}$ and $C_{\mathrm{surf}}$,
all in parallel, and that combination in series with $C_{\mathrm{var}}$:
\begin{equation}
    C_{e} = \left(\frac{1}{2(C_{\mathrm{gap}}+C_{\mathrm{surf}})} + 
        \frac{1}{C_{\mathrm{var}}}\right)^{-1}
    \label{eq.elecap}
\end{equation}
Because there are many capacitances in parallel, 
changes in $C_{var}$ lead to a wider frequency shift than 
analogous varactor tuning of the magnetic side.  
Using more precise equation\hyp{}based analysis (available in 
\Cref{s:unit_cell_analysis}) that matches our qualitative analysis, we 
cross\hyp{}check and finalize design parameters
$R$, $g$, and $w$ for the magnetic and electric meta-atoms.

In \Cref{f:theory:freq}, we observe that the difference 
in resonant frequencies between 0 and 10~V for the electric meta-atom are larger 
than the magnetic meta-atom.  Hence, since the effect of 
$C_{\mathrm{var}}$ differs, 
overlapping of resonance will not always occur when $U_{M}=U_{E}$.
Rather than of simply finding the area where $U_{M}=U_{E}$
as suggested above, 
we instead need to search for the voltage pair for every desired 
phase that also maximizes the
reflection or transmission magnitude. 
We do this by running one-time optimization that searches for the 
voltage pair that maximizes $|T|$ (or $|\Gamma|$) for each phase 
and generates a static lookup table that will later be used for beam steering.

\begin{figure}[t]
\centering
Biasing lines---failed attempts \textbf{(a)--(c):}\\
\begin{subfigure}[b]{.33\linewidth}
\centering
\includegraphics[width=.9\linewidth]{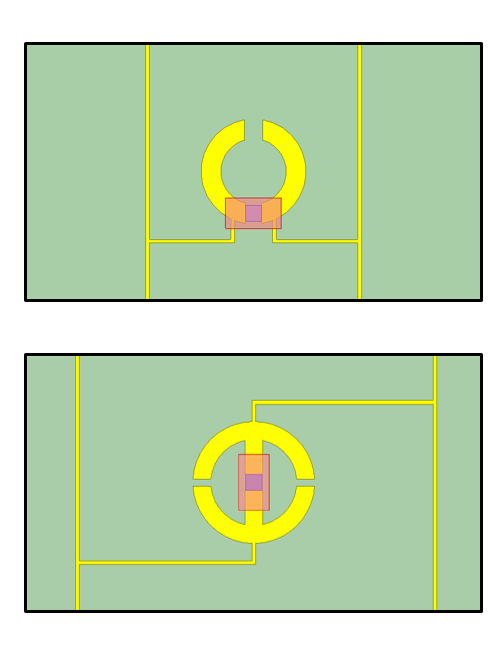}
\caption{Copper line}
\label{f:copperline}
\end{subfigure}
\begin{subfigure}[b]{.33\linewidth}
\centering
\includegraphics[width=.9\linewidth]{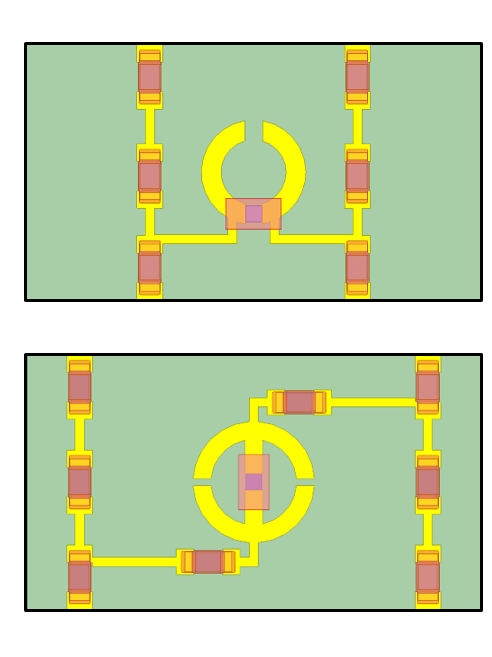}
\caption{Coil inductor}
\label{f:coilinductor}
\end{subfigure}
\begin{subfigure}[b]{.32\linewidth}
\centering
\includegraphics[width=.9\linewidth]{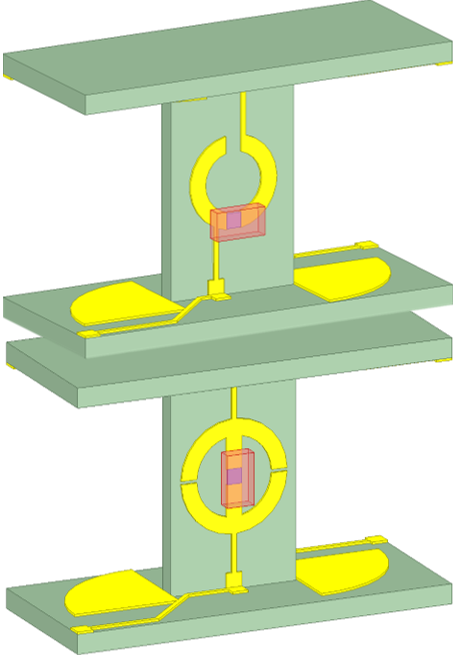}
\caption{Radial stub}
\label{f:radialstub}
\end{subfigure}
\begin{subfigure}[a.]{1\linewidth}
\centering
\includegraphics[width=\linewidth]{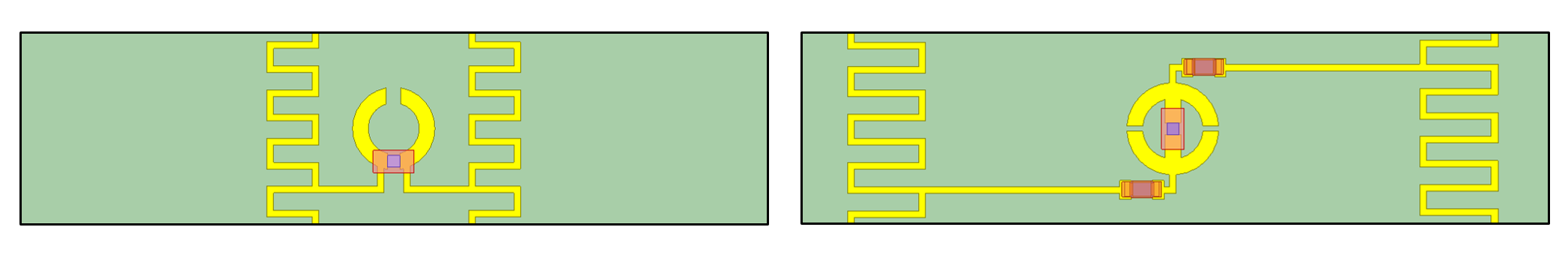}
\caption{\textbf{\shortnames{}} \emph{meander} biasing line design.}
\label{f:meander_m}
\end{subfigure}
\caption{Biasing line designs: notable failed attempts include
\textbf{(a)}~straight microstrip, \textbf{(b)}~coil inductor, and 
\textbf{(c)}~radial stub. \shortname{} uses 
an inner meander line \textbf{(d)} for magnetic, and 
an outer meander line \textbf{(e)} for electric 
meta\hyp{}atoms.}
\label{f:biasing}
\end{figure}

\subsection{Control Network Design}
\label{s:design:control}

To control the meta-atoms, we connect an off\hyp{}surface control unit
via ribbon cables with on\hyp{}surface \emph{biasing lines}, which altogether 
comprise the entire \emph{control network}
(\Cref{f:high_level_design} 
on p.~\pageref{f:high_level_design}).

\subsubsection{Biasing lines}
\label{s:design:control:line}

This design process concerns 
the problem of designing the
on\hyp{}surface control network to interact
with mmWave\hyp{}frequency meta\hyp{}atoms.
Directly connecting a line to the meta-atoms changes the performance of the meta-atom, which
causes mmWave signal loss and
invalidates the design process described 
previously (\S\ref{s:design:unit}).
To mitigate such adverse effects, we seek to
design biasing lines that incorporate
radio frequency (RF) chokes, low pass filters that 
block RF signals within a certain frequency band from
propagating on direct current (DC) signal paths. 
Our primary design goals are to design a biasing network that \textbf{1)}~minimizes 
the use of extra components, \textbf{2)}~avoids a large amount of 
copper on the panel where the meta-atom is placed, and 
\textbf{3)}~is straightforward to fabricate. 
This is challenging because 
mmWave meta\hyp{}atoms are sensitive to the shape and 
placement of the choke. 

\parahead{Failed attempts.}
\Cref{f:biasing} shows various biasing line structures we have considered. 
First, we try a straight copper line design \textbf{(a)}. We use a 
narrow width resembling a very high impedance transmission 
line, to try to attenuate the RF signal while the DC biasing voltage is 
applied. However, to achieve the desired impedance, a very 
narrow width transmission line (0.07~mm) is required which is not 
possible to fabricate by common PCB manufacturing techniques. 

Second, we try the use of inductors to create a high\hyp{}impedance line
\textbf{(b)}.  The impedance of an inductor is determined by the RF
frequency and is proportional to its inductance. However, inductance of mmWave inductor components are limited. Hence, we would need to apply at least four inductors in series to achieve the desired isolation, introducing significant
surface complexity and also internal resistance that adversely affects
unit cell efficiency.

Third, a radial stub which is an open ended transmission line is employed. The
length of the stub determines the input impedance of the line, and so thus acts as an RF ``choke''
that blocks mmWave signals, while a DC biasing voltage is applied to the cell
from the control network.
The required length of the stub is one\hyp{}quarter wavelength, which is comparable 
to the cell size. But if the stub is designed on the same panel, the stub itself 
would reflect most of the wave, stealing energy to illuminate the cell itself. 
To avoid this problem, one can put the stubs on a perpendicular panel, as shown in
\Cref{f:radialstub}. This could potentially solve the wave reflection issue, 
but would complicate
implementation, since there would be one perpendicular panel for each unit cell.

\parahead{Proposed \textit{meander} structure.}
To achieve our design goals, we have formulated a meander 
structure that acts as an RF choke, but at the same time 
connects the vertically adjacent meta-atoms. 
Longer and thinner traces provide more inductance, so
by bending the straight wire vertically and horizontally,
we enable the control network itself to be
an inductor that outperforms the multiple off-the-shelf inductors.
But this increases capacitance between the two meander lines
on opposing sides of the unit cell, which also invalidates
our meta\hyp{}atom design process.  So \shortname{}
places the meander line of the magnetic meta-atom in a non-overlapping 
configuration relative to the meander line of the electric meta-atom.
To compensate the loss from the microstrip that connects the electric 
meta-atom and the meander lines,
we add two off-the-shelf inductors next to the electric meta-atom.
The final result is shown in \Cref{f:meander_m}.

\begin{figure*}[ht]
    \begin{subfigure}[b]{.24\linewidth}
    \centering
    \includegraphics[width=1\linewidth]{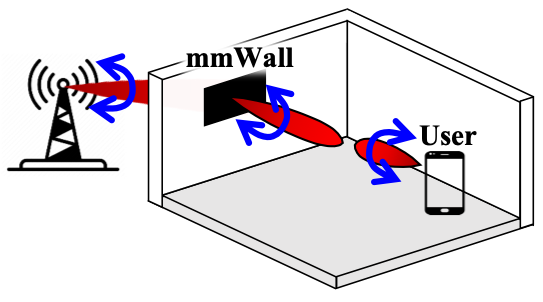}
    \caption{Refractive establishment}
    \label{f:ll1}
    \end{subfigure}
    \hfill
    \begin{subfigure}[b]{.24\linewidth}
    \centering
    \includegraphics[width=1\linewidth]{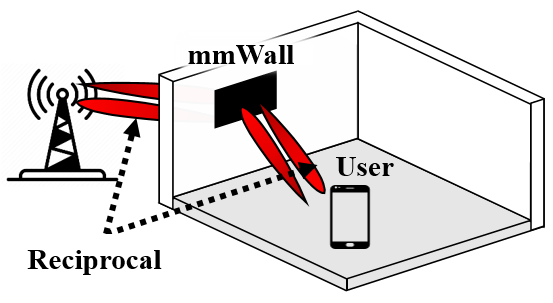}
    \caption{Angular reciprocity}
    \label{f:ll4}
    \end{subfigure}
    \hfill
    \begin{subfigure}[b]{.24\linewidth}
    \centering
    \includegraphics[width=1\linewidth]{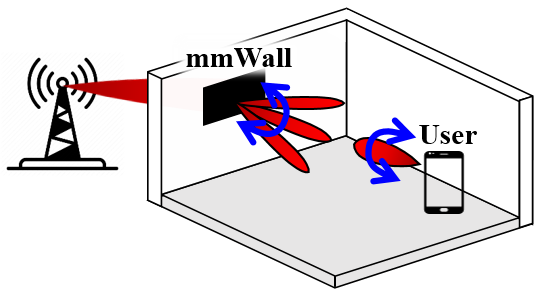}
    \caption{Beam tracking}
    \label{f:ll3}
    \end{subfigure}
    \begin{subfigure}[b]{.24\linewidth}
    \centering
    \includegraphics[width=1\linewidth]{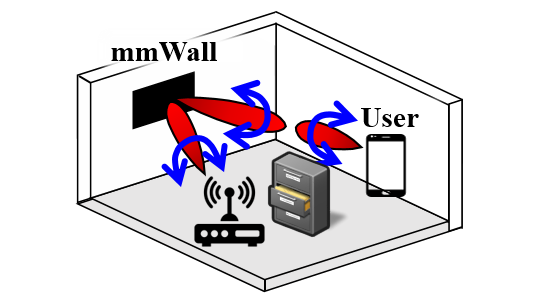}
    \caption{Reflective establishment}
    \label{f:link_reflective_establishment}
    \end{subfigure}
    \caption{\shortnames{} refractive link establishment, 
    angular reciprocity property,
    tracking, and reflective link establishment.}
    \label{f:align}
\end{figure*}

\subsubsection{Beam steering and splitting}
\label{s:design:control:steer}

A conventional phased array transmitter has a net radiation pattern
multiplying the radiation pattern of a single element by
the \emph{array factor} (AF), the pattern induced by the array.
Unlike prior mmWave receive\hyp{}transmit relay systems which
require two phased antenna arrays (one to receive 
and another to transmit a new phase\hyp{}shifted signal), 
\shortname{} uses only a single array of meta-atoms to
directly shift the phase of an existing mmWave signal. 
For $L$ omni antennas with $d$ separation,
each with transmit amplitude $A$, 
$\mathrm{AF} = A\sum_{n=0}^{L-1} e^{2\pi jnd(\cos\theta)/\lambda}$
with radio wavelength $\lambda$ and steering angle $\theta$. 

\shortname{} applies different phase shifts to each 
meta\hyp{}atom group for beam steering. Specifically, by
searching over the space of control voltages to maximize
reflection or transmission amplitude subject to achieving
the desired phase (\Cref{f:huygen_heatmap}), 
we construct a look\hyp{}up table
that maps steering phase $\varphi$ to the chosen
unit cell voltage pair (and without loss of generality)
transmission coefficient: 
$\Phi(\varphi) \rightarrow \langle U_M, U_E,\Gamma\rangle$.
The difference with conventional beamforming is that element
amplitudes vary, so \systemnames{} net radiation pattern becomes 
$\sum_{n=0}^{L-1}\Phi_\Gamma(\phi)e^{j\phi}$
where $\phi = 2\pi n d \cos\theta$. 


To transform a single beam into multi-armed beams,
we modify the above AF to account for 
angles $\theta_{1}$ and $\theta_{2}$: 
\begin{equation}
\sum_{n=0}^{N-1} \left(\alpha \Phi_{\Gamma}(\phi_1) 
    e^{j\phi_1} + 
    \beta \Phi_{\Gamma}(\phi_2) e^{j\phi_2}\right)
\label{eq:array_factor_hms}
\end{equation}
where $\phi_k = 2\pi n d \cos\theta_k$, and $\alpha$ and $\beta$ are 
weighting terms that that determine the power of each beam.

\subsection{Link Layer Design} 
\label{s:design:link}

Recall that \shortname{} operates in two different modes,
a lens mode and a reflective mode.
\textbf{1)}~In lens mode,
a mmWave signal refracts through \shortname{} 
allowing, \emph{e.g.}, a user inside the building 
to communicate with the base station (\emph{ENodeB})
in a cellular network.
This requires two beam alignments: 
one between the ENodeB and \shortname{}, and 
another between \shortname{} and the user. 
\textbf{2)}~In mirror mode, \shortname{} reflects mmWave signals.
For example, in wireless LAN settings, 
it reflects the beam between the AP and user, 
which requires beam alignment between the AP and \shortname{},
and again between \shortname{} and the user.

\shortname{} electronically 
switches between the two modes because
different users may be located outdoors and indoors. 
Hence, \shortname{} sweeps the beam in both lens and mirror mode 
to align to the user during a beam search.

\parabreak{}Our development here follows the outline of the 
existing 5G New Radio (NR) beam management protocol, but adapts 
it to \systemnames{} unique capabilities.
The current 5G NR beam search proceeds
in three steps: 
\textbf{1)}~the ENodeB sweeps its beam, the user
equipment (\emph{UE})
selects a best direction,
and reports it to the ENodeB; 
\textbf{2)}~the ENodeB refines its beam (\emph{i.e.}, sweeping a
narrower beam over a narrower range),
the user detects the best direction and reports it to the ENodeB; 
\textbf{3)}~the ENodeB fixes a beam and the 
UE refines its receiver beam.

To establish a link from a cold start,
the ENodeB sweeps different directions such that 
the user can detect the best beam for an initial link establishment
(\Cref{f:ll1}). 
If the UE cannot detect the beam or the beam strength is low, 
it turns \shortname{} to a lens mode and signal
it to simultaneously sweep the beam received from the ENodeB, via 
sub\hyp{}6~GHz control.
At the same time, the UE scans its receiving beam to various directions.
After the search, the UE knows the combination of the ENodeB's 
transmit beam angle, 
\shortnames{} beam refraction angle, 
and its receive beam angle that maximizes the SNR of downlink signals.
Given an initial link, ENodeB and \shortname{} refine the beam 
by simultaneously sweeping narrower beams over narrower ranges,
and lastly, the user refines its receiving 
beam.\footnote{We note that some full\hyp{}duplex relays 
\cite{abari2017enabling} 
require the relay node's receive direction aligned to the ENodeB,
which is not necessary with \shortname{}.}
ENodeB\hyp{}\shortname{} alignment takes $O(n)$ steps 
(for $n$ directions), 
and \shortname{}\hyp{}UE alignment takes $O(n^2)$ steps, so
cold\hyp{}start beam alignment as described above
takes $O(n^3)$ steps, but 
only once \emph{ever} when \shortname{} is installed,
because both ENodeB and \shortname{} are stationary. 
Hence, the common case of 
cold\hyp{}start beam establishment between \shortname{} and user 
in fact requires $O(n^2)$ steps (\emph{cf.} \Cref{f:ll3}).
Also, the above notably does not require modifications
to the existing 5G NR protocols.

As illustrated in \Cref{f:ll4} and demonstrated 
experimentally in \Cref{s:eval:microbenchmarks}, \shortname{}
refracts beams in one direction at the same angle
as they arrive at the surface from the other side of
the surface (angular reciprocity), which obviates
the need for separate downlink and uplink link establishment. 

Since the UE controls \shortname{}, 
the user can alternate between the `lens' mode for 
outdoor\hyp{}to\hyp{}indoor communication 
and the `mirror' mode for indoor communication.
For example, when the user switches from an outdoor to an indoor
ENodeB, it signals \shortname{} to re\hyp{}establish the beam estimation 
process for indoor usage.

\parahead{Multi-beam search.}
\shortname{} can create irregular beam shapes such as multi\hyp{}arm 
beams (\Cref{s:eval:e2e}), which allows it to leverage 
state\hyp{}of\hyp{}the\hyp{}art beam searching algorithms that exploit
the sparsity of the mmWave channel to accelerate beam search
\cite{10.1145/3005745.3005766, 10.1145/3032970.3032974} by orders
of magnitude improvement (essential for agile and mobile applications 
such as VR), now for the first time at a surface.

\begin{figure*}
    \begin{subfigure}[b]{.261\linewidth}
    \centering
    \includegraphics[width=1\linewidth]{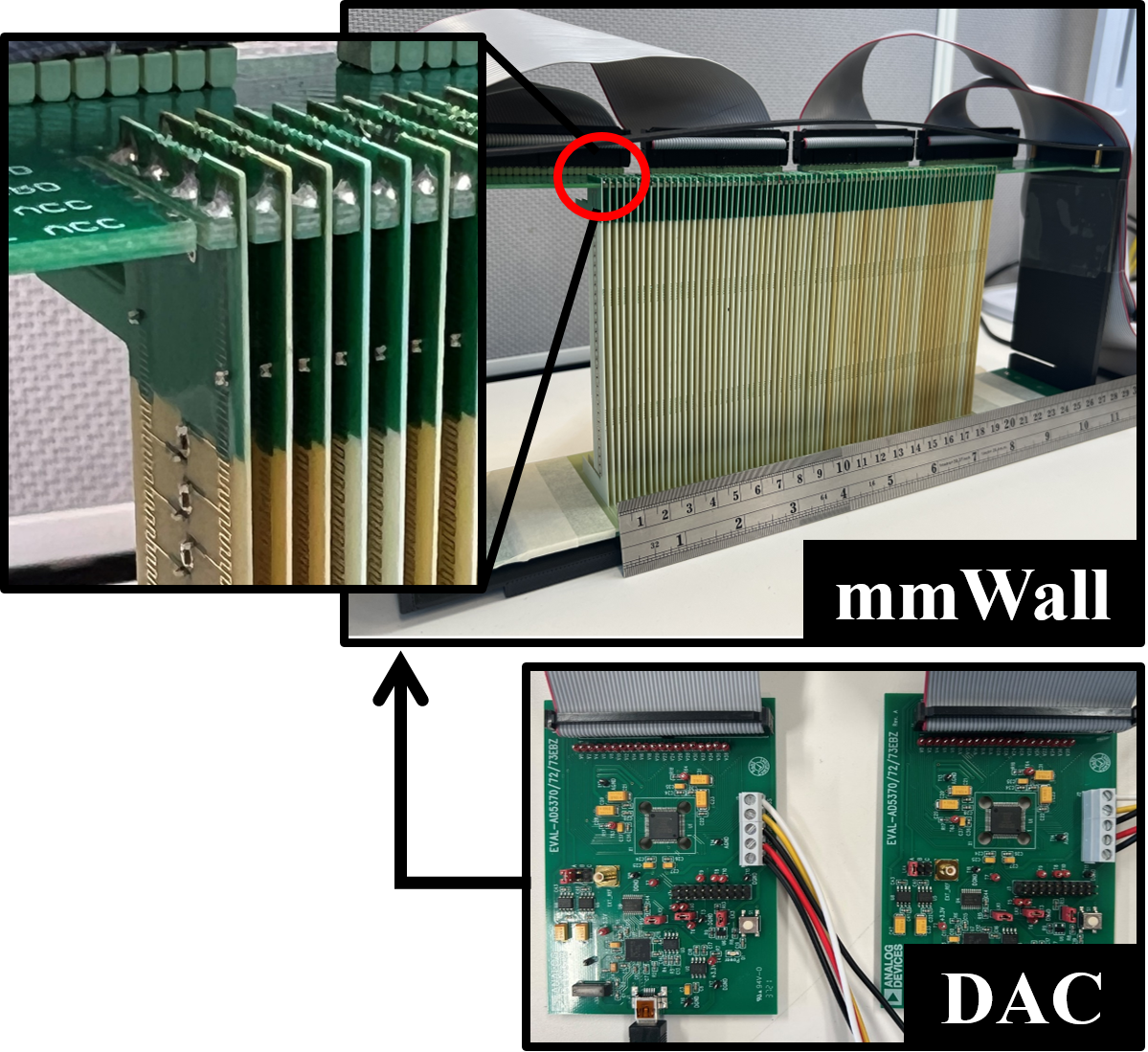}
    \caption{\shortname{} hardware}
    \label{f:impl:mmwall_close}
    \end{subfigure}
    \begin{subfigure}[b]{.194\linewidth}
    \centering
    \includegraphics[width=1\linewidth]{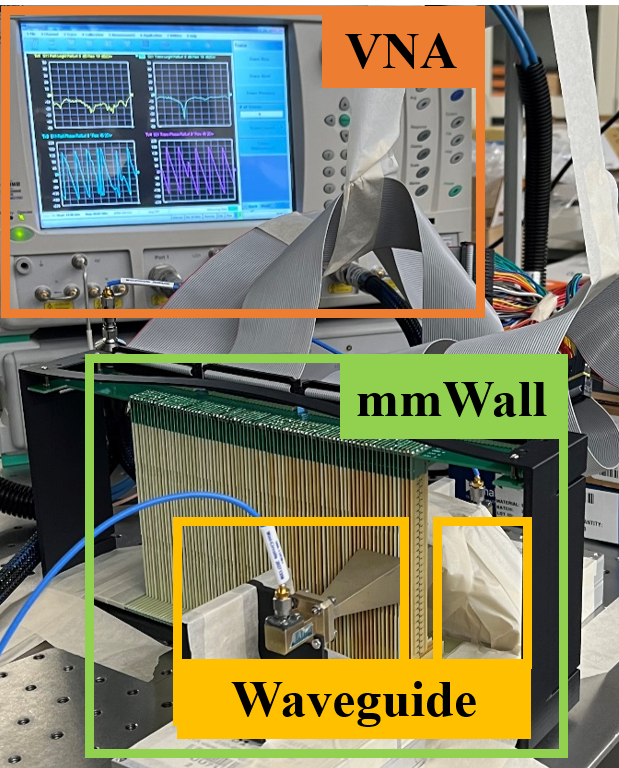}
    \caption{Near-field testing}
    \label{f:impl:nearfield}
    \end{subfigure}
    \begin{subfigure}[b]{.255\linewidth}
    \centering
    \includegraphics[width=1\linewidth]{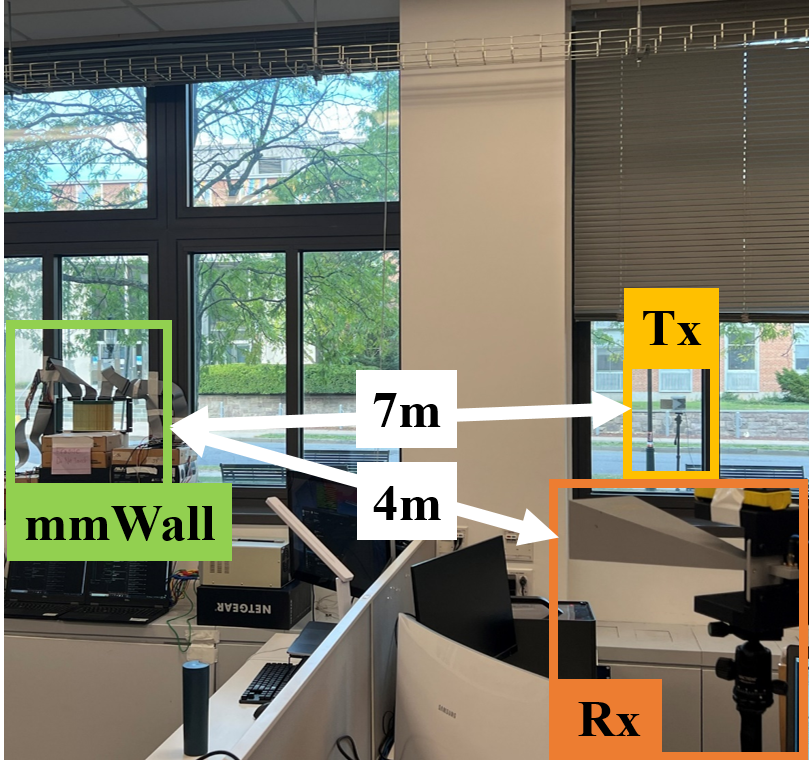}
    \caption{Outdoor\hyp{}to\hyp{}indoor scenario}
    \label{f:eval:scenario:out2in}
    \end{subfigure}
    \begin{subfigure}[b]{.275\linewidth}
    \centering
    \includegraphics[width=1\linewidth]{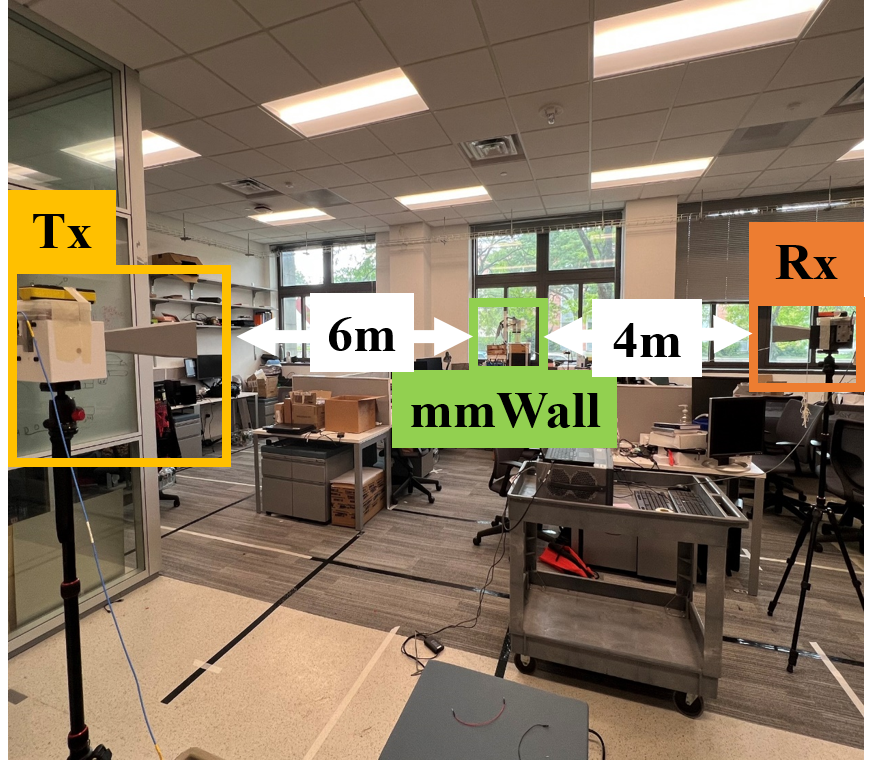}
    \caption{Indoor reflective scenario}
    \label{f:eval:scenario:indoor}
    \end{subfigure}
    \caption{\shortnames{} hardware implementation, 
    transmissive (`lens') and indoor reflective (`mirror') evaluation
    scenarios.}
    \label{f:impl}
\end{figure*}

\section{Implementation}
\label{s:impl}

\begin{figure}
\centering
\includegraphics[width=1\linewidth]{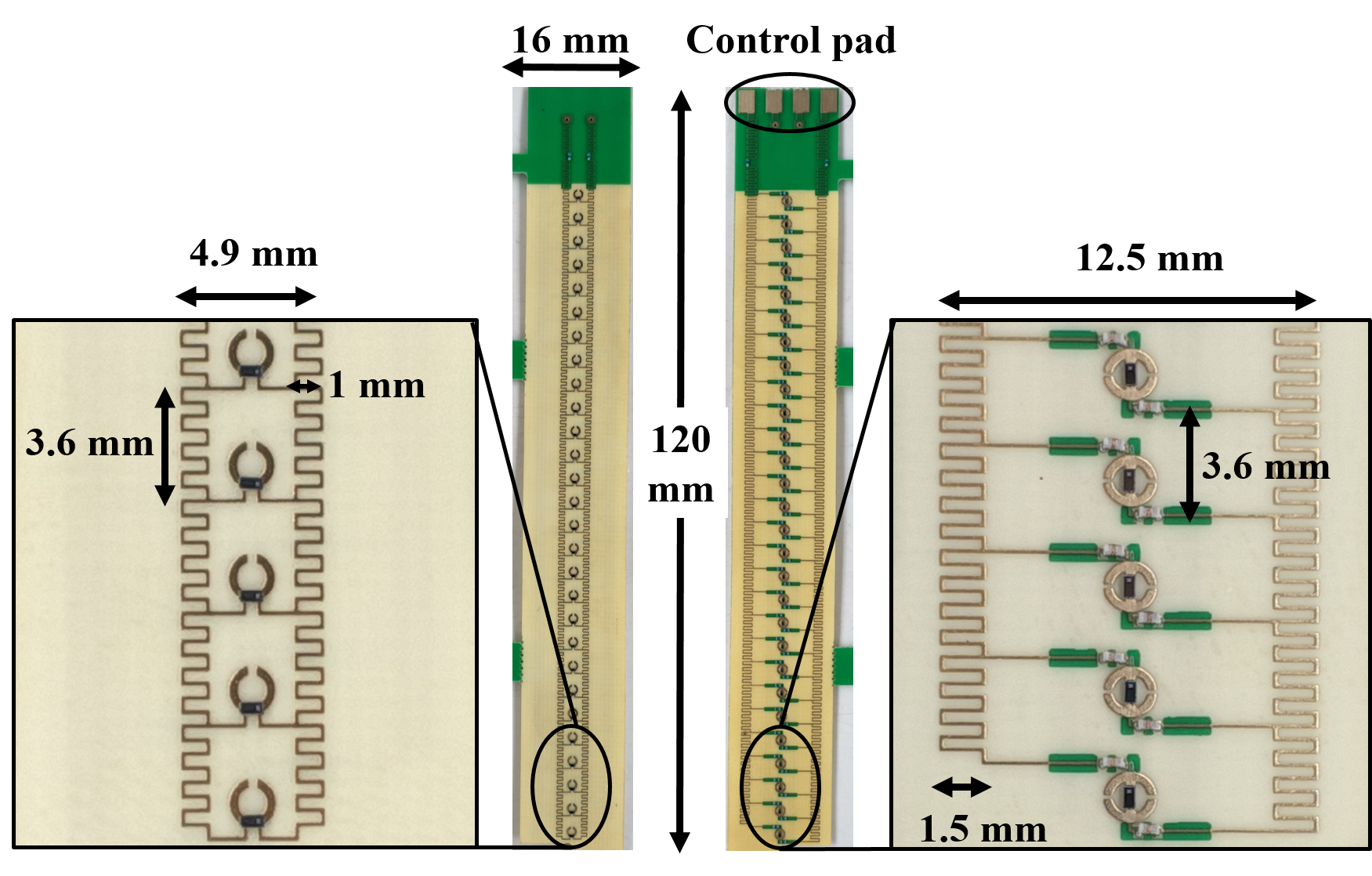}
\caption{\shortnames{} \emph{ribs}, comprised of our
proposed meta\hyp{}atom design fabricated on a Rogers
printed circuit board.}
\label{f:ribs}
\end{figure}

We have fabricated and assembled a complete hardware prototype of \shortname{},
summarized in \Cref{f:impl}.
\shortnames{} meta\hyp{}atoms are fabricated on a 16 by 120~mm \emph{rib}
made of Rogers 4003C printed circuit board (PCB) substrate, as shown in \cref{f:ribs}.
We assemble the PCB and constituent Macom 
\href{https://www.macom.com/products/product-detail/MAVR-000120-14110P}{MAVR-000120-1411} 
varactor diodes\footnote{We have modeled this varactor based on its 
\emph{Simulation Program with Integrated
Circuit Emphasis} (SPICE) model (see Appendix, \Cref{f:cvar_v_control}).} 
and Coilcraft \href{https://www.coilcraft.com/en-us/products/rf/ceramic-core-chip-inductors/0201-(0603)/026011c/026011c-1n7/}{026011C-1N7} inductors. 

\begin{figure}
\centering
\includegraphics[width=.9\linewidth]{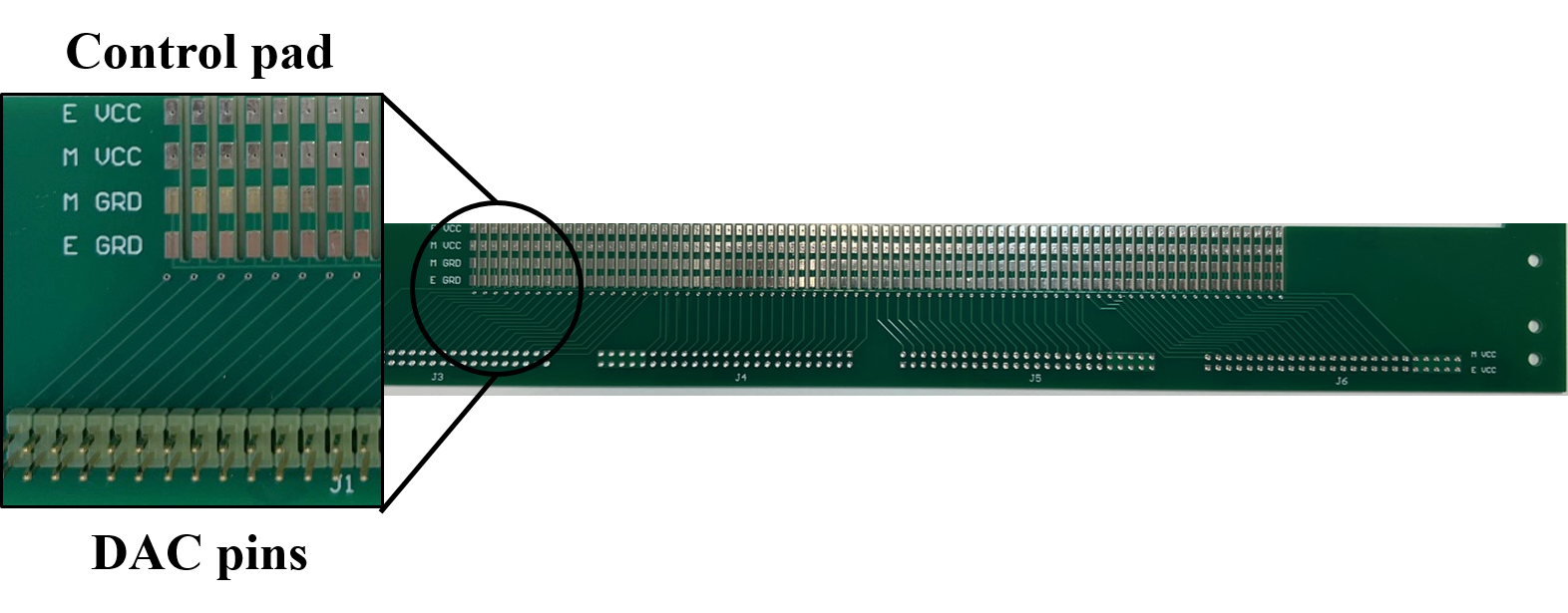}
\caption{\shortnames{} FR4 holder\fshyp{}control board.}
\label{f:holder}
\end{figure}

In total, we have fabricated $76$ ribs, each consisting of $28$ vertical meta-atoms. These ribs are mechanically hold together with two perpendicular FR4 panels; one in top and the other in bottom of the structure. The top FR4 also provides control lines as it is shown in \cref{f:holder}. Each rib's control pads are then soldered to the upper holder board, which connects 
the ribs to a DAC through its microstrip traces and pin headers.
The lower holder boards are installed to position and the ribs fixed
into these boards.
For holding the ribs and Fr4 panels steady, a 3D printed enclosure is fabricated that 
provides a standing support, as shown in \Cref{f:impl:mmwall_close}.
The spacing between the adjacent ribs are $2.6$ mm, making 
the dimension of our \shortname{} prototype $120\times197.6$~mm.
We note that scaling up our prototype with identical ribs and expanded
FR4 holder boards is straightforward.

Four 40-channel 
\href{https://www.analog.com/en/design-center/evaluation-hardware-and-software/evaluation-boards-kits/eval-ad5370.html}{AD5370} 
16-bit DACs from Analog Devices
allow independent control of both electric and magnetic cells of every \shortname{} rib. 
Each DAC supplies a variable 0 to 10~V control voltage for each of 40 
channels (\textit{i.e.}, one DAC per $20$ boards with one channel for
$U_E$ and $U_{M}$ apiece). We have programmed the DACs in Microsoft Visual 
C++ along with a voltage look-up table for each transmissive 
or reflective steering angle, or multi\hyp{}arm beam combination.




\section{Evaluation}
\label{s:eval}

\begin{figure*}
    \begin{subfigure}[b]{.275\linewidth}
    \centering
    \includegraphics[width=1\linewidth]{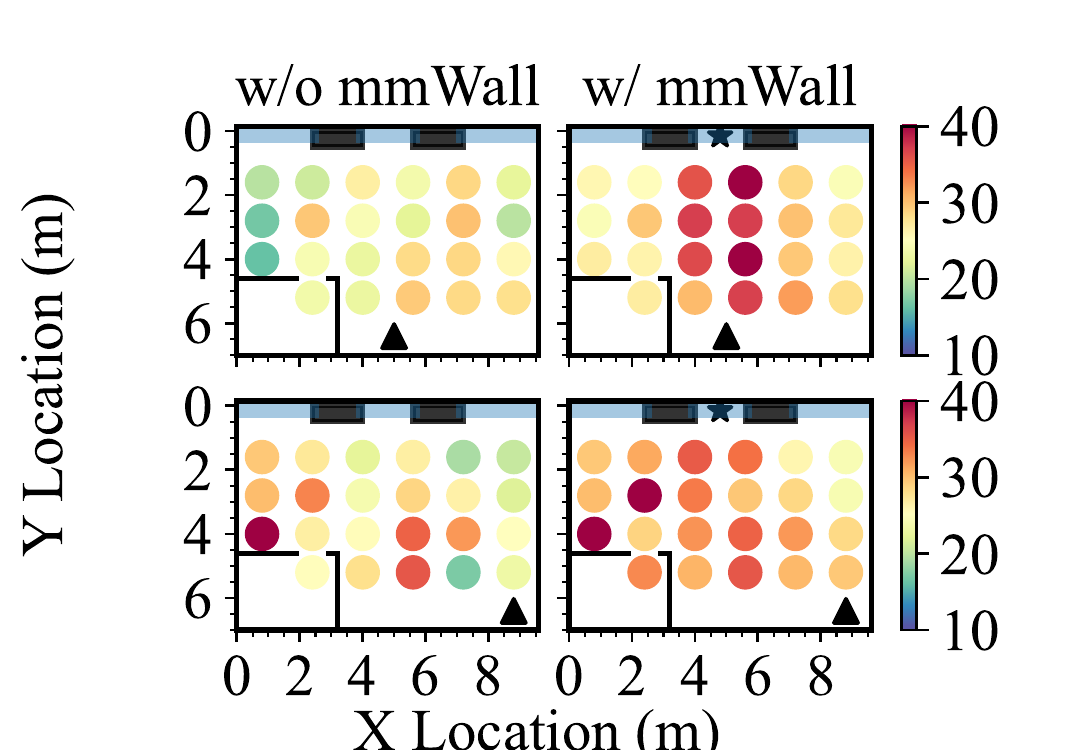}
    \caption{Indoor-to-indoor.}
    \label{f:eval:map:indoor}
    \end{subfigure}
    \begin{subfigure}[b]{.45\linewidth}
    \centering
    \includegraphics[width=1\linewidth]{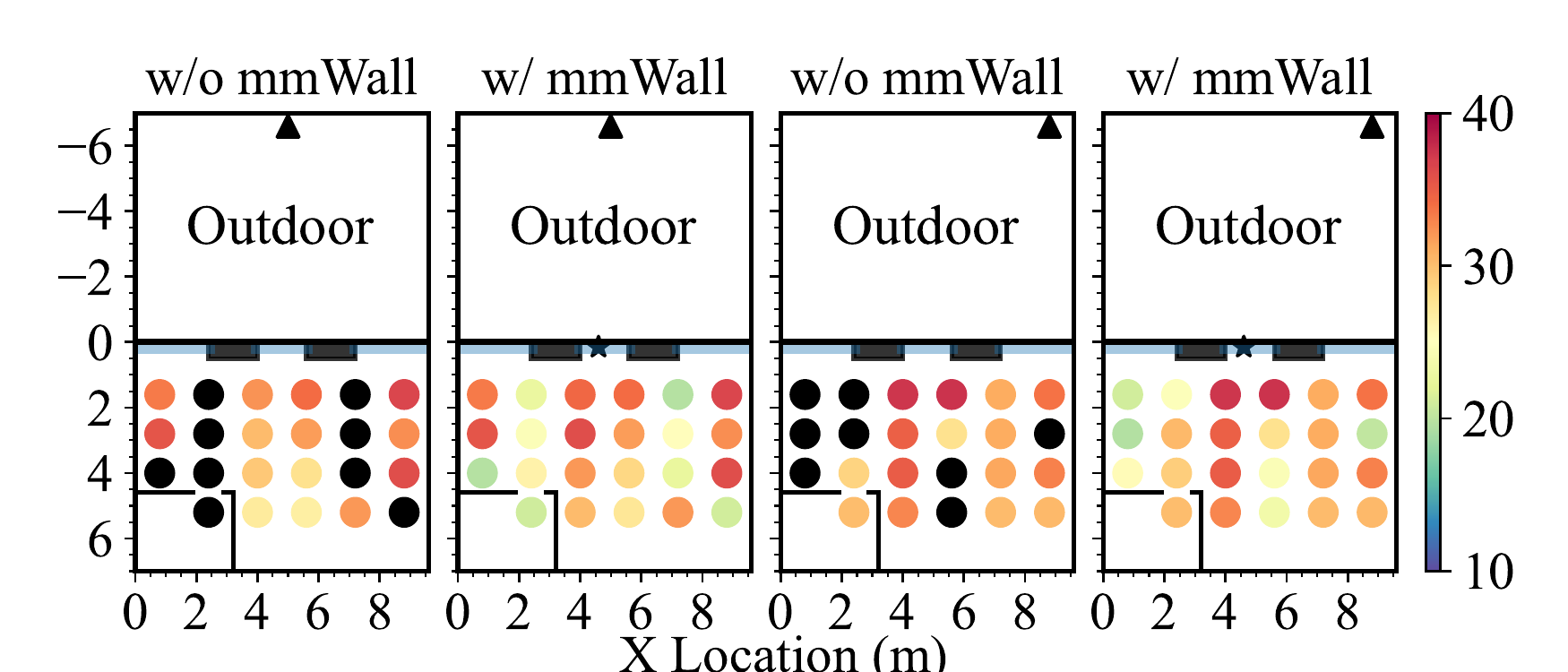}
    \caption{Outdoor-to-indoor.}
    \label{f:eval:map:outdoor}
    \end{subfigure}
    \begin{subfigure}[b]{.265\linewidth}
    \centering
    \includegraphics[width=.98\linewidth]{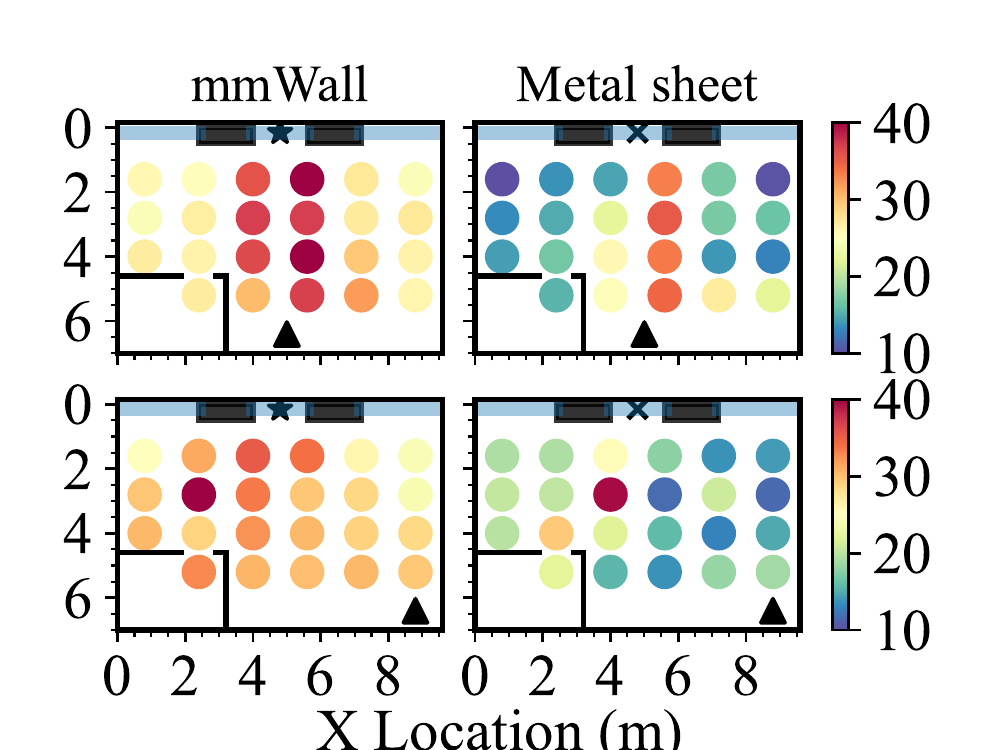}
    \caption{Comparing to a metal sheet}
    \label{f:eval:map:comparison}
    \end{subfigure}
    \caption{\shortname{}'s SNR improvement over the best NLoS environment path when (a) both transmitter and receiver are located indoor (\textit{upper}: transmitter facing \shortname{} in perpendicular direction; \textit{lower}: transmitter facing $30^{\circ}$ away from \shortname{})  and (b) when transmitter is located outdoor (\textit{left}: transmitting in perpendicular direction; \textit{right}: transmitting in $30^{\circ}$ off-angle). We use the following notations: \shortname{} $\scriptstyle\bigstar$, transmitter $\blacktriangle$, receiver $\bigcirc$. $\CIRCLE$ indicates no signal.}
    \label{f:eval:map}
\end{figure*}

We begin with field studies that quantify \shortname{}'s SNR gain compared to the best NLoS environment path for both indoor-to-indoor and outdoor-to-indoor links (\S\ref{s:eval:e2e}). 
Moreover, we explore the SNR gain and link failure rate under dynamic link conditions. 
We then evaluate multi-armed beams created by \shortname{} at various receiver locations (\S\ref{s:eval:multiarm}).
We conclude with microbenchmarks to characterize \shortname{}'s steering performance, its support for wide steering angle, angular reciprocity, operation across wide bandwidths, and the impact of the surface size (\S\ref{s:eval:microbenchmarks}). 

\subsection{Methodology}
\label{s:eval:methodology}

We evaluate in various indoor and outdoor scenarios.
In indoor-to-indoor settings, 
both receiver and transmitter are located in a $10\times8$ m office 
with interior walls, windows, and a server room. 
In between the three windows, there are two brick walls (black rectangles in \cref{f:eval:map}).
For outdoor-to-indoor testbed, the receiver is inside the office 
while the transmitter (denoted as triangle in \cref{f:eval:map}) is located outdoors ($6-7$~m away from the exterior window).
During the experiments, we place \shortname{} in front of the window inside the room. 
The loss of window is approximately $-4$ to $-5$ dB.
For each outdoor-to-indoor and indoor-to-indoor experiment,
we conduct two sets of experiments, each with  
one fixed transmitter location and $23$ receiver locations. 
The transmitter is $6.3$~m away from \shortname{} 
and faces the surface perpendicularly in the first set. 
The second set has the transmitter module $6.8$~m away from \shortname{},
and its beam hits the surface with approximately $30^{\circ}$ to $40^{\circ}$ angle. 
For every set, $23$ different receiver locations are identical.
During the beam search, \shortname{} steers the angle by the step of $0.5^{\circ}$.
For end-to-end performance, we report SNR with a noise floor of $80$~dBm.

\parahead{Near-field experiments.}
To accurately measure a Huygens pattern and create a 
voltage-to-phase look-up table,
we collect near-field reflection and transmission coefficients of \shortname{}
using two-port Anritsu MS4647B VNA, operating from 70~kHz to 70~GHz, as shown in \cref{f:impl:nearfield}.
The Huygens pattern measured from the VNA is shown in \cref{s:vna}.
To minimize measurement error, we perform a two port calibration before acquiring the data.
For data collection, we program the VNA using LabVIEW, which communicates with four DACs through the socket.
During the measurement, \shortname{} is placed in between two waveguide horn antennas that are connected to the VNA.
Since the area of \shortname{} is larger than the aperture of waveguide horns, 
we collect the pattern on multiple locations of \shortname{}. 
In \cref{s:vna}, we present a measured Huygens pattern at different locations of \shortname{}
and demonstrate the robustness of \shortname{} against fabrication variations.

\parahead{Far-field experiments.}
With a 25~dBi transmit horn antenna, our calculated EIRP is 31~dBm.
We use the same antenna at receiver 
but apply a $-10$~dB correction to reflect
typical UE antenna gain.
To generate mmWave signals, we use off-the-shelf phase-locked loop (PLL) frequency synthesizers \href{https://www.analog.com/en/design-center/evaluation-hardware-and-software/evaluation-boards-kits/eval-adf4371.html#eb-overview}{ADF4371} with integrated VCO and frequency quadrupler,
which quadruples $6.125$ GHz VCO signals to $24.5$ GHz. 
At transmitter, since the PLL output power is  $<-13$~dBm, we use the PLL 
in conjunction with a variable gain amplifier (VGA)
\href{https://www.analog.com/media/en/technical-documentation/data-sheets/hmc997g.pdf}{HMC997LC4}, which amplifies signals by 18 to 20~dBm.  

\subsection{In-situ Performance}
\label{s:eval:e2e}
In this section, we evaluate the end-to-end performance of \shortname{} for indoor and outdoor scenarios.

\begin{figure}[t]
    \centering
    \includegraphics[width=1\linewidth]{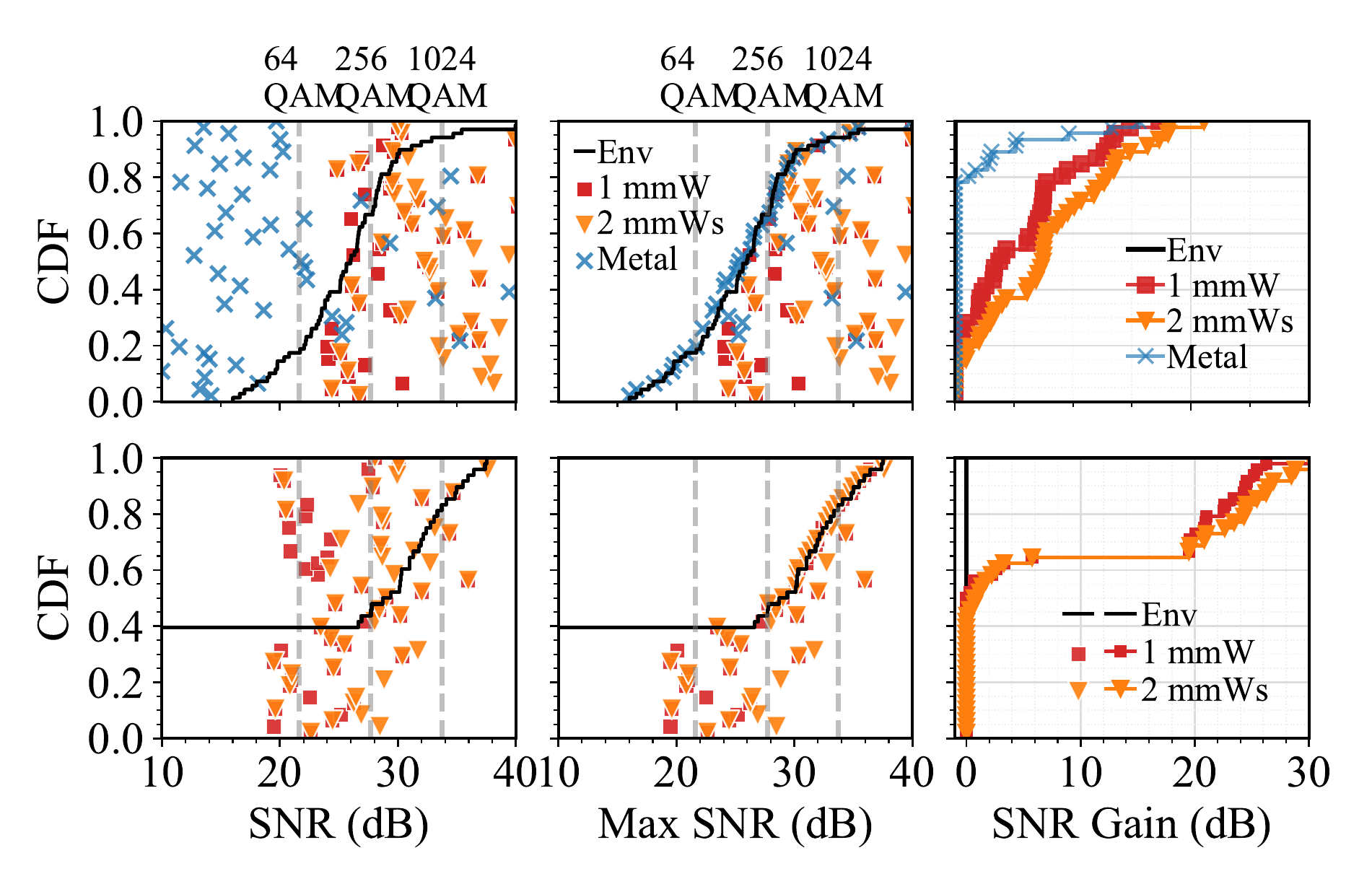}
    \caption{SNR gain of one or more \shortnamepl{} over various Rx locations (\textit{upper}: indoor-to-indoor; \textit{lower}: outdoor-to-indoor scenarios). The left subfigures show the SNRs of both the best NLoS environment path and one or more \shortname{} paths at the corresponding Rx location. The middle subfigures shows the maximum SNR with or without one or more \shortnamepl{}. The right subfigures shows the actual SNR gain of one or more \shortnamepl{} against the best environment path across various Rx locations.}
    \label{f:eval:cdfs}
\end{figure}

\parahead{SNR improvement over the best environment path.}
We first evaluate SNR measurements without and with \shortname{} 
at multiple transmitter and receiver locations 
(two locations for the transmitter and $23$ locations for the receiver)
when a LoS path is blocked. 
For each link, the transmitter and receiver modules (and \shortname{} if deployed) search for a NLoS path 
that maximizes SNR.
In \cref{f:eval:map}, we demonstrate the measurements before and after deploying \shortname{}. 
First, \cref{f:eval:map:indoor} reports
SNRs with the transmitter facing the window at $0^{\circ}$ (upper subfigure) and $30^{\circ}$ (lower subfigure) in the indoor testbed. 
Both subfigures show that our indoor testbed has a rich scattering environment,
providing SNR above $25$ dB for some receiver locations.
However, the receivers at the left or right corner of the room often get SNR below $20$ dB.
With \shortname{}, all receivers, including the ones in the corner, achieve SNRs of at least $24$~dB. 
Also, the nodes located within \shortname{}'s $-45^{\circ}$ to $45^{\circ}$ steering angle get SNRs greater than $30$ dB. 
\shortname{}'s SNR improvement is more evident in \cref{f:eval:cdfs}.
The left subfigure of \cref{f:eval:cdfs} evaluates a CDF of 
best environment SNRs (denoted by black curves)
along with SNR of \shortname{} links at the corresponding receiver location (denoted by markers). 
The middle subfigure shows the CDFs of maximum SNRs between the environment and \shortname{} links.
The right subfigure shows CDFs of the SNR gains per each receiver location.
For the indoor-to-indoor testbed (upper subfigures), 
\shortname{} guarantees 91\% of locations outage-free under 128-QAM~\cite{nishimori2012interference} mmWave data rates 
while only 40-50\% receivers achieve 128-QAM in the absence of \shortname{}
Moreover, among 80\% of the receivers who experience the gain from \shortname{}, 
some receive more than 15~dB SNR boost.
In \cref{f:eval:map:outdoor}, we evaluate SNR improvement in outdoor-to-indoor scenarios.
Without \shortname{}, the receivers who cannot establish a NLoS link through the window 
suffer from complete link failure. 
With \shortname{}, on the other hand, all receivers get SNRs of at least $19-20$ dB.
The lower subfigures in \cref{f:eval:cdfs} demonstrate the CDFs of outdoor-to-indoor SNR improvement.
A single \shortname{} guarantees 64-QAM for almost all Rx locations.  
Also, it boosts SNR by up to 30~dB for $40\%$ of the links.  
Our results reveal that \shortname{} is very beneficial when mmWave signals suffer from the wall blockage.

\parahead{Deploying multiple \shortname{}s.}
To evaluate more than one \shortname{}, 
we place another \shortname{} in front of the window on the right side of the room. 
\cref{f:eval:cdfs} demonstrates the SNR gain 
from deploying two \shortnamepl{} for indoor-to-indoor links (upper subfigures). 
Compared to the gain from a single \shortname{}, 
two provide $\le 5$~dB SNR gain for some links. 
For outdoor-to-in links (lower subfigures), there is almost no gain from adding an extra \shortname{}.
The results indicate that a single \shortname{} is sufficient to provide a good coverage (at least $128$-QAM for reflective and $64$-QAM for transmissive links) in a $10\times 8$ m office room. 
In a static environment another \shortname{} will not help
\emph{if} a \shortname{} path is already available. 

\begin{figure}[t]
    \centering
    \includegraphics[width=.98\linewidth]{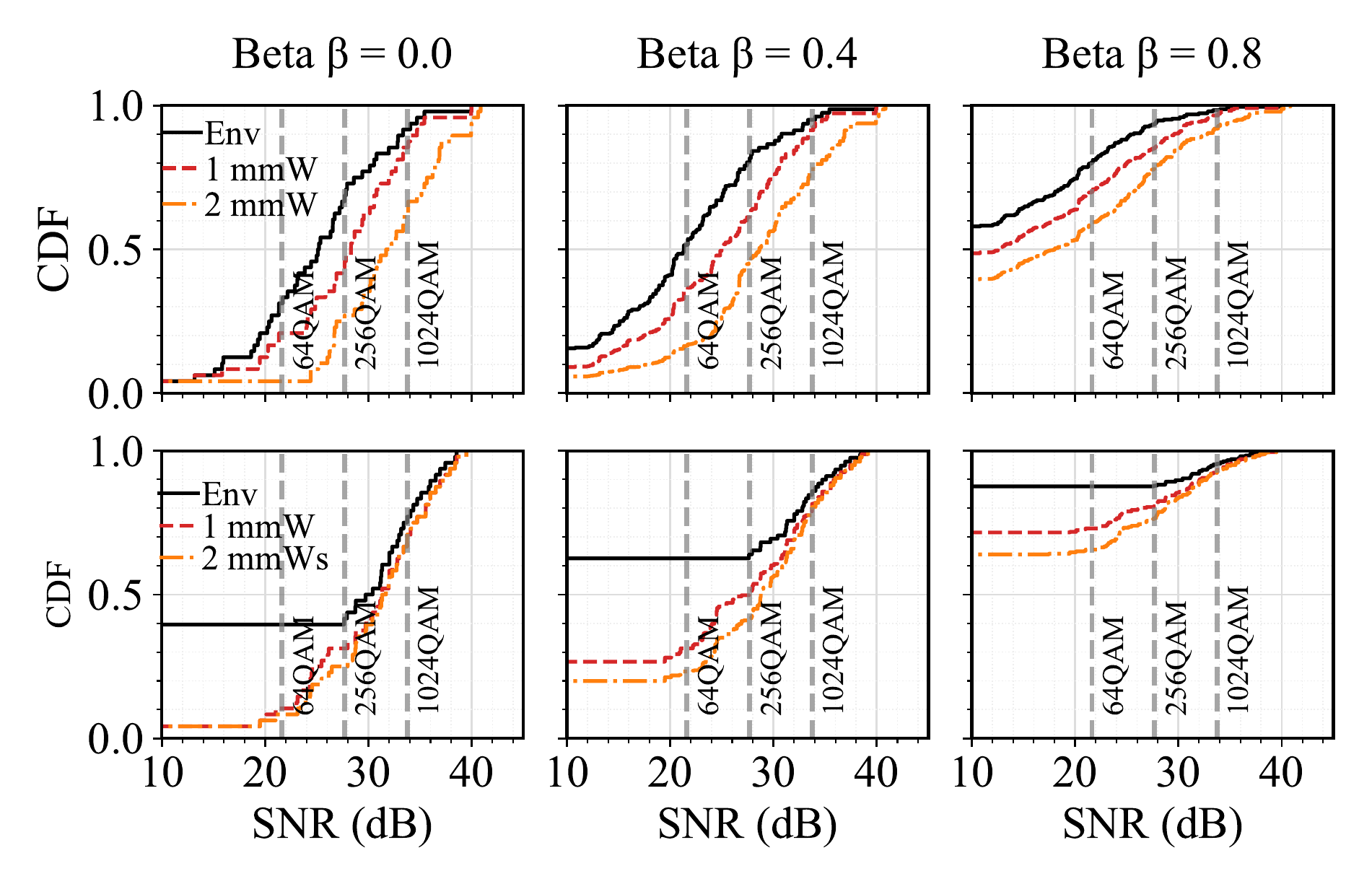}
    \caption{SNR improvement (multiple \shortnamepl{})
    for dynamic links (\textit{upper}: indoor-to-indoor; \textit{lower}: outdoor-to-indoor scenarios). $\beta$ is probability of environment path or surface blockage.}
    \label{f:eval:dynamic}
\end{figure}

\parahead{Improving reliability for dynamic links.}
While a single \shortname{} delivers good SNRs across all receiver locations, 
blockage can always occur on \shortname{} links. 
Similarly, even if a strong NLoS environment path exists,
this link can always be blocked.
At mmWave frequencies, the indoor environment commonly provides 
three to four strong paths, including the LoS path~\cite{9514544}. 
Since the number of available paths is limited, 
an increase in the number of blockages can easily lead 
to link failure, which exacerbates when these blockages start to move.
The key advantage of deploying one or more \shortname{}s 
lies in the improvement in link reliability.
By providing a diverse, strong alternative path, \shortname{} reduces the probability of link scarcity. 
In \cref{f:eval:dynamic}, we demonstrate the SNR gain across various Rx locations 
as the \textit{blockage probability} for both environment and \shortname{} links $\beta$ increases.
In indoor-to-indoor scenarios, a single \shortname{} and two \shortname{}s reduce the probability of link failure by a ratio of up to 10\% and 20\% under 80\% path blockage, respectively. 
For the outdoor testbed, the probability of link failure decreases by 40\% for a single \shortname{} and 45\% for two \shortname{}s under 40\% blockage probability. 
Hence, we conclude that multiple \shortname{}s are beneficial when channel environments are highly dynamic.


One may argue that deploying a simple reflecting metal sheet could 
help, but \shortname{}'s ability to steer the beam makes a huge difference in coverage.
We evaluate SNRs of links reflected by a $60\times60$~cm metal sheet 
along with SNRs of links steered by $10\times20$~cm \shortname{}.
As shown in \cref{f:eval:map:comparison}, only 10\% of all receivers achieve SNRs above 30~dB, 
and the rest receive SNRs below 15~dB. 
Also, it is worth noting that for a metal sheet, SNRs depend a lot on the transmitter location.
In \cref{f:eval:map:comparison}, only 4\% achieve good SNR when the transmitter is located in the corner. 
Moreover, in \cref{f:eval:cdfs}, only 8\% of all receivers achieves more than 5~dB SNR gain from the metal sheet. 
On the other hand, \shortname{} guarantees at least 25~dB SNRs across all areas.  
We conclude that compared to a fixed-angle reflection, \shortname{} links are 
less sensitive to transmitter, receiver, and surface location, 
making them more robust.

\begin{figure}[t]
    \begin{subfigure}[b]{1\linewidth}
    \centering
    \includegraphics[width=.9\linewidth]{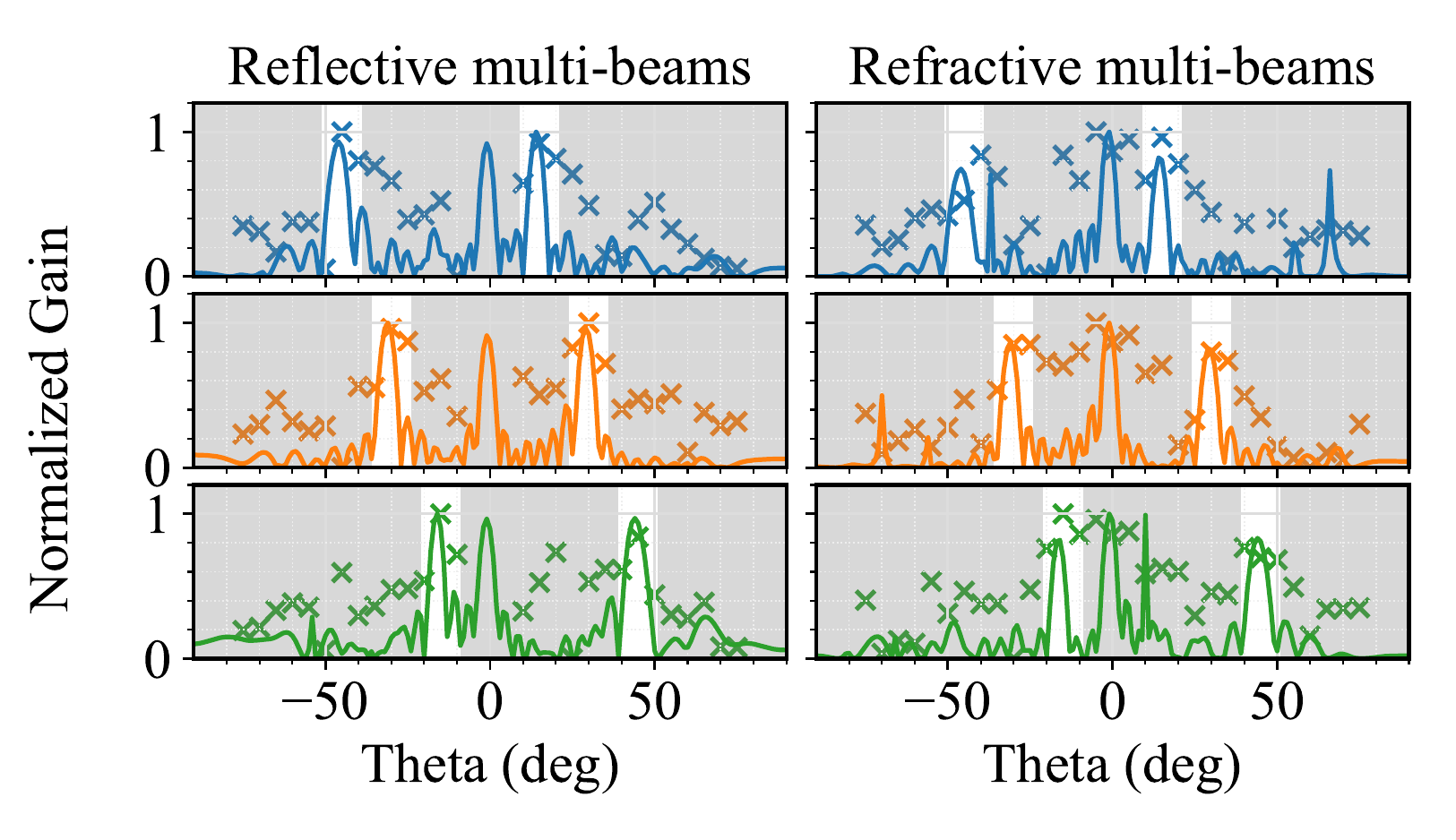}
    \caption{\shortnames{} multi\hyp{}armed beam pattern (\textit{upper:} 
    $-45$/$15^{\circ}$ degree beam split; 
    \textit{middle:} $-30$\fshyp{}$30^{\circ}$ split;
    \textit{lower:} $-15$\fshyp{}$45^{\circ}$ split). Empirical 
    points are denoted $\times$, with simulation curves.}
    \label{f:eval:multibeam_pattern}
    \end{subfigure}
    \begin{subfigure}[b]{1\linewidth}
    \centering
    \includegraphics[width=.9\linewidth]{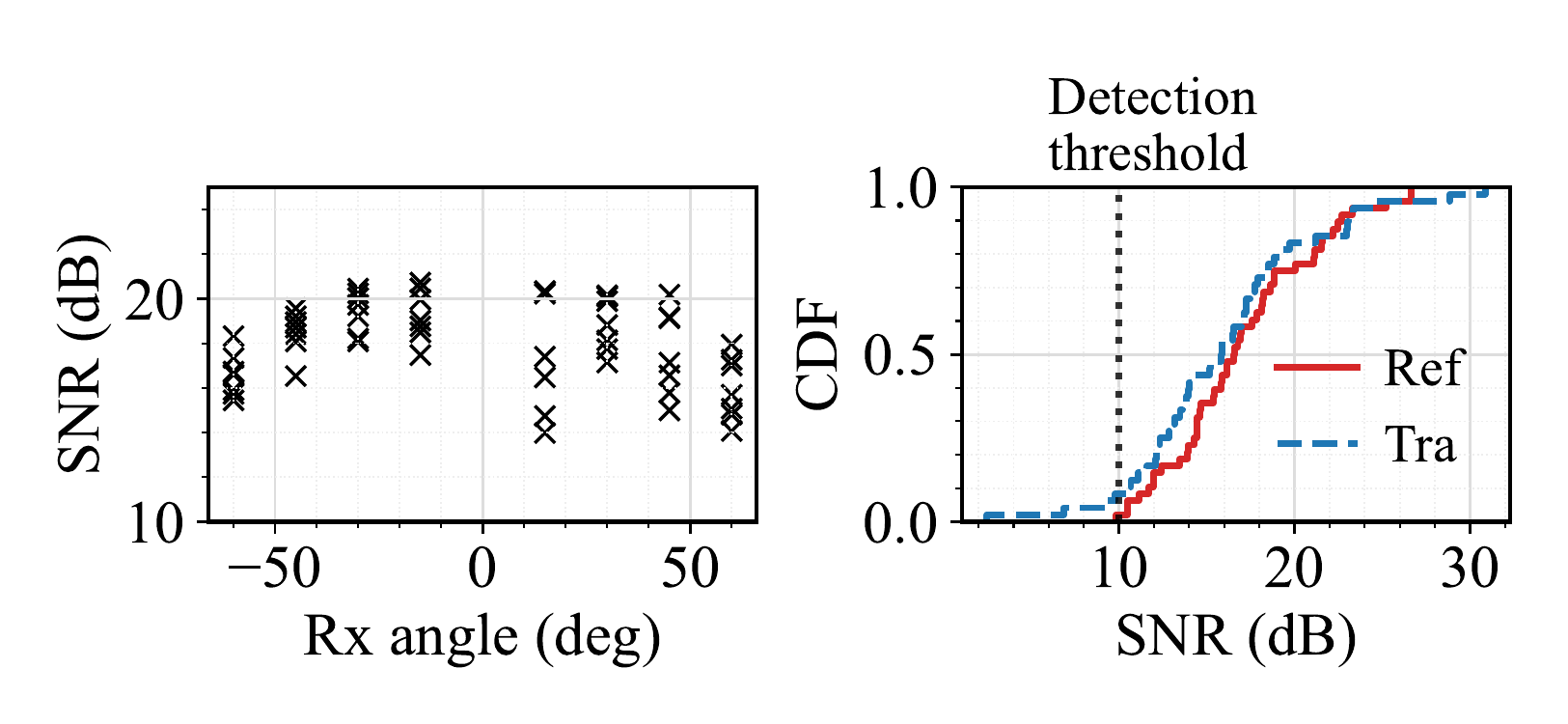}
    \caption{SNR of aligned multi-beams (\textit{left:} a fixed distance between Tx and \shortname{} and between Rx and \shortname{}; \textit{right:} various Tx and Rx locations in the office setting.)}
    \label{f:eval:multibeam_detection}
    \end{subfigure}
    \caption{Evaluation of \shortname{}'s multi-armed beams.}
    \label{f:eval:multibeam}
\end{figure}

\begin{figure*}[ht]
    \centering
    \includegraphics[width=1\linewidth]{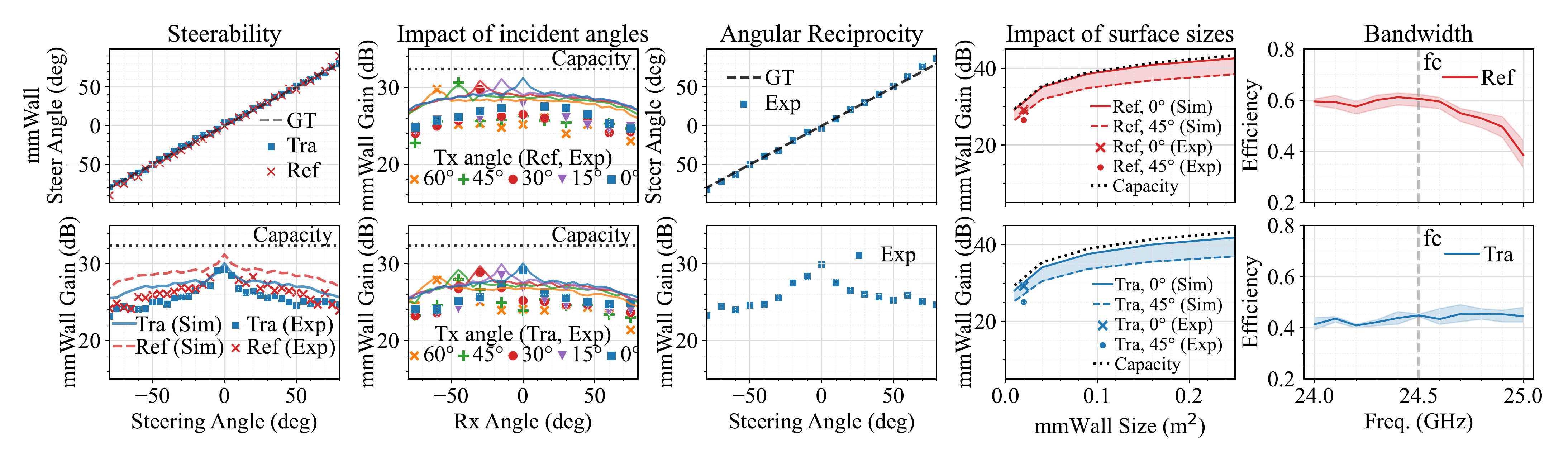}
    \caption{Microbenchmarks evaluating (\emph{left} to \emph{right}:)
    surface steerability, performance sensitivity of incident 
    mmWave angle, angular reciprocity, surface size, and frequency bandwidth. Empirical 
    points are denoted with markers, with simulation curves.}
    \label{f:microbenchmark}
\end{figure*}




\subsection{Multi-armed Beams}
\label{s:eval:multiarm}

We next evaluate \shortname{}'s capability to generate multi-armed beams.
In \cref{f:eval:multibeam_pattern}, we present our measurements on the multi-armed beams with simulation results.
Specifically, \shortname{} splits an incident beam into two beams at $-45^{\circ}$/$15^{\circ}$ and steers the multi-beams to $-30^{\circ}$/$30^{\circ}$ and $-15^{\circ}$/$45^{\circ}$. 
To measure the beam pattern, we locate the transmitter and receiver three meters away from \shortname{} and record the gain of \shortname{} as we move the receiver from $-90^{\circ}$ to $90^{\circ}$ angle with respect to \shortname{}.
Since we did not measure the beam pattern in an anechoic chamber, the received beam interfered with signals reflected off of the indoor environment.
Despite the interference, we observe that the gain peaks at the angles
where \shortname{} splits the beam. 
Furthermore, as \shortname{} steers its multi-armed beams, the measured peak changes accordingly.
It is worth noting that transmissive multi-beams show a peak at $0^{\circ}$ due to leakage from the transmitter directly to the receiver.
We then measure SNRs as \shortname{} generates and steers various multibeams (\textit{i.e.} beams that are $15^{\circ}$ to $120^{\circ}$ apart from each other).
The distance between the transmitter and \shortname{} and 
between the receiver and \shortname{} are fixed to $2$ m.
The left subfigure of \cref{f:eval:multibeam_detection} reveals 
that as the beam is splitted into a wider angle, SNR drops. 
 
To demonstrate the feasibility of a multi-armed beam search, 
\shortname{} again splits the beam into two beams that are $15^{\circ}$ to $120^{\circ}$ apart from each other. 
Then it aligns the beam with the receivers at $23$ different locations in the room.
In \cref{f:eval:multibeam_detection}, we report SNRs of \shortname{}'s multi-beam links aligned with various receivers. 
The results show that more than $90\%$ of multi-beam links achieve SNRs above $10$~dB.
Considering that no signal is detected in many locations for outdoor-to-indoor settings, 
$10$~dB SNR is enough for the receiver to detect the beam and start the alignment.
We conclude that \shortname{} can generate multi-beam whose signal is strong enough to accelerate beam search.

\subsection{Microbenchmarks}
\label{s:eval:microbenchmarks}

We now evaluate \shortname{}'s steering performance, 
its support for wide steering angle, angular reciprocity, 
operation across wide bandwidths, and the impact of the surface size.
The microbenchmark testbed consists of the receiver and 
transmitter module that are three meters away from \shortname{}. 
We present both the experimental measurements and simulated results acquired from HFSS simulation.

Since \shortname{} does not have an amplifier, 
the most determinant factor of its gain is the effective aperture $A_{e}$.
A well-defined relation for the effective aperture in terms of the aperture gain
$G$ is $A_{e}4\pi/\lambda^2$.
We define the aperture gain as our capacity and compare it
against our measured \shortname{} gains throughout microbenchmarks.
A rigorous analysis on \shortname{} gain is available in \cref{s:path_loss_model}.


\parahead{\shortname{} controllability.}
In \cref{f:microbenchmark}, we evaluate the \shortname{} controllability.
First, we present \shortname{}'s beam alignment accuracy in the upper subfigure. 
We place the receiver at $37$ locations in our testbed and estimate the angle that provides maximum SNR as \shortname{} sweeps the beam from $-90^{\circ}$ to $90^{\circ}$ angle.
During the experiment, the transmitter is facing \shortname{} at $0^{\circ}$ angle.
For both reflection and transmission, \shortname{} accurately steers the beam with at most $3^{\circ}$ difference with the groundtruth (GT).  
Second, we evaluate the effect of steering angle on the \shortname{} gain in the lower subfigure. 
As \shortname{} increases the steering angle, \shortname{} gain decreases.
Furthermore, reflection provides slightly better gain than transmission. 
\begin{figure}[t]
    \begin{subfigure}[b]{1\linewidth}
    \centering
    \includegraphics[width=.95\linewidth]{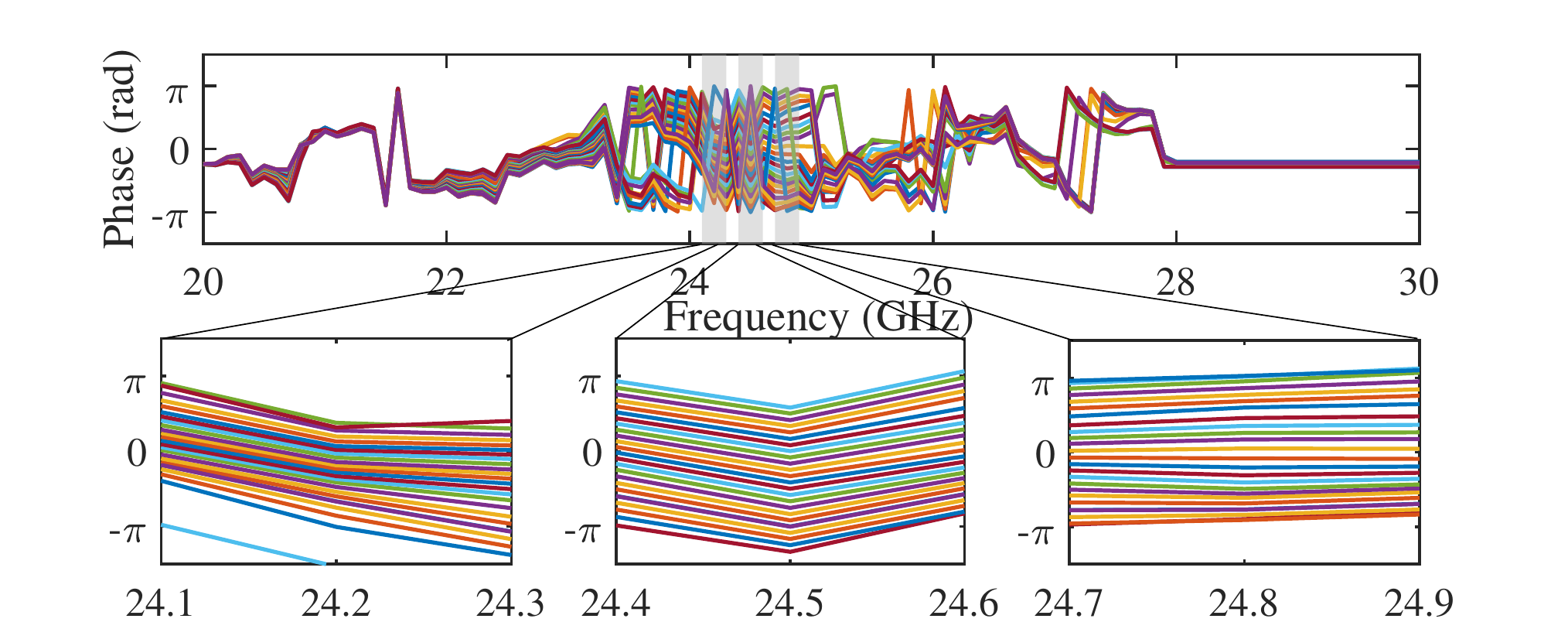}
    \caption{Reflection $\Gamma$}
    \label{f:eval:bwd_ref}
    \end{subfigure}
    \begin{subfigure}[b]{1\linewidth}
    \centering
    \includegraphics[width=.95\linewidth]{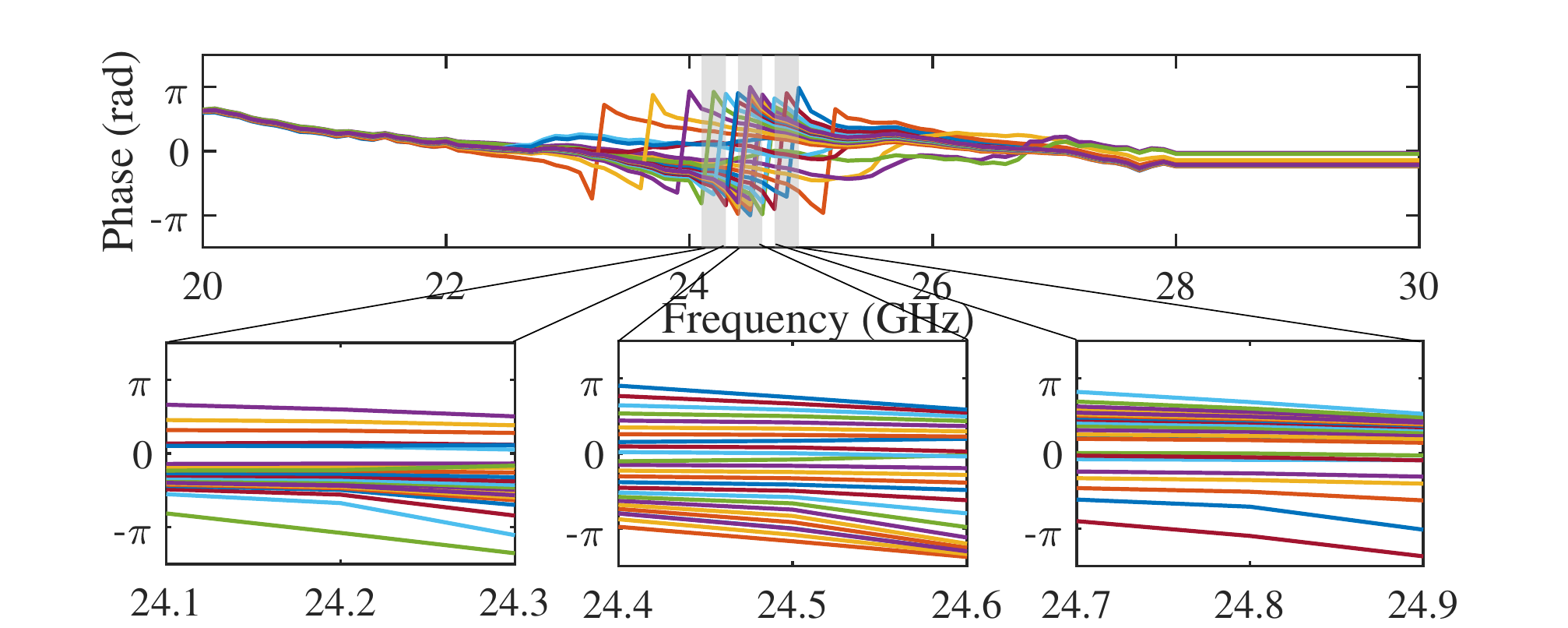}
    \caption{Transmission $T$}
    \label{f:eval:bwd_tra}
    \end{subfigure}
    \caption{\shortname{}'s phase coverage and consistency 
    (VNA measurement) across different frequencies. 
    The curves indicate the voltage pairs ($U_{M}$, $U_{E}$) that provide $-180^{\circ}$ to $180^{\circ}$ phase with the step of $15^\circ$ at $24.5$ GHz. The unwrapped phases across \shortname{}'s operating bandwidth is shown.}
    \label{f:eval:phase_freq}
\end{figure}

\parahead{Support for wide steering angle.}
In this microbenchmark, we move the transmitter to $5$ different locations and the receiver to $37$ locations.
By doing so, we evaluate the effect of an incident beam angle on the \shortname{} gain jointly with the steering angle.
For both reflection (upper subfigure) and transmission (lower subfigure), 
increasing the incident beam angle does not greatly reduce the \shortname{} gain. 
Important observation is that even with $135^{\circ}$ steering angle (for example, Tx angle at $60^{\circ}$ and Rx angle at $-75^{\circ}$), \shortname{} achieves more than $22$ dB gain, indicating that \shortname{} is capable of refracting the beam in a very wide angle. 


\parahead{Angular reciprocity.}
Once \shortname{} aligns the downlink channel, uplink is also aligned due to its angular reciprocity.
To demonstrate this property, we evaluate the uplink beam alignment accuracy and 
the corresponding \shortname{} gains when the downlink alignment is already established.
The upper subfigure of \cref{f:microbenchmark} demonstrates that the uplink beam alignment is very accurate and is within an error of $3^{\circ}$. Also, the \shortname{} gain is above $23$ dB for all steering angles.

\parahead{Operation across wide bandwidths.}
To demonstrate \shortname{}'s phase coverage across wide bandwidth, 
we present our VNA measurements from $20$ to $30$~GHz.
In \cref{f:eval:phase_freq}, each curve indicates 
the phase response of a voltage pair in the lookup table 
we found at our center frequency, $24.5$~GHz. 
Here, we emphasize three points.
First, \shortname{} provides a full phase coverage from $-\pi$ to $\pi$
over the 200~MHz 5G mmWave link bandwidth. 
Second, within 200~MHz bands (highlighted in gray), phase distributions are 
mostly constant, allowing improvements over the entirety of those bandwidths.
Third, \shortname{} can operate in the entire $23.5$ to $25.5$~GHz band,
as it provides a wide range of phase there.
Hence, \shortname{} operates over the mmWave 5G bandwidth.
Yet, our goals in designing the meta-atoms are 
to reduce transmission or reflection \textit{loss level} with full phase coverage. 
To quantify both magnitude and phase coverage at the same time, 
we define \textit{efficiency} as
$\sum^{180}_{\phi=-180} (Te^{-1j\phi})/360$ where $T$ is a set of points obtained from near-field transmissive (or reflective) huygens pattern that provides a maximum magnitude for $-180^{\circ}$ to $180^{\circ}$ phases.
\cref{f:microbenchmark} demonstrates that for both reflection and transmission, 
the efficiency is consistent and declines after $24.9$~GHz.
Since targeted operational 
bandwidth for \href{https://www.qualcomm.com/news/releases/2021/07/qualcomm-completes-worlds-first-5g-mmwave-data-connection-support-200-mhz}{5G mmWave} is 200~MHz, we conclude that \shortname{} operates within the 5G bandwidth.

\parahead{Increasing \shortname{} size.}
In \cref{f:microbenchmark}, we simulated gain increase as \shortname{} size increases from $10\times 10$ cm to $50\times 50$ cm with $0^{\circ}$ steering 
(for reflection, it is a specular reflection) and $45^{\circ}$ steering. 
Also, we compare our simulated results with the effective aperture-based capacity.
\shortname{} gain at both $0^{\circ}$ and $45^{\circ}$ steering 
increases with increasing surface 
size, following the capacity trend.

\section{Conclusion}
\label{s:concl}

This paper presents \shortname{}, the first Huygens metasurface that can 
reconfigures  itself to relay an incoming mmWave beam as either a
non\hyp{}specular ``lens'' or ``mirror.''  
Our prototype steers single\hyp{} or multi-armed beams 
at non\hyp{}specular directions, arbitrarily
in real-time.
We conduct an extensive evaluation 
in various indoor and outdoor settings, demonstrating
significant SNR improvement, and describe how 
scaling to even larger sizes is eminently possible.

\section{Acknowledgements}
This work is supported by the National Science Foundation under grant CNS-1617161, Natural Sciences and Engineering Research Council of Canada (NSERC), Canada Foundation for Innovation (CFI) and Ontario Research Fund (ORF).
\let\oldbibliography\thebibliography
\renewcommand{\thebibliography}[1]{%
  \oldbibliography{#1}%
  \setlength{\itemsep}{0pt}%
}
\clearpage
\bibliographystyle{concise2} 
\bibliography{paper}
\clearpage
\appendix
\section{Unit Cell Electromagnetic Analysis}
\label{s:theoretical_analysis}
\label{s:unit_cell_analysis}

We now present a full mathematical analysis 
of \shortnames{} unit cells.
Since electromagnetic fields are naturally continuous 
and will not change the propagation characteristics by itself, 
we artificially introduce electric and magnetic surface 
currents $(\vec{J}_{s}, \vec{M}_{s})$
from the electric and magnetic meta-atoms, enforcing a field discontinuity:
\begin{equation}
    \begin{aligned}
        \vec{J_s}=\hat{n}\times [H_{t}-H_{i}],~\vec{M_s}=-\hat{n}\times [E_{t}-E_{i}]
    \end{aligned}
\end{equation}
where $\hat{n}$ is a unit normal.
The average tangential field applied 
on the meta-atom pair induces ($\vec{J_{s}},\vec{M_{s}}$).
To induce suitable surface currents,
we need a proper surface impedance for each meta-atom:
\begin{equation}
    \begin{aligned}
        \hat{n}\times [E_{avg}]=Z_{e}\vec{J_{s}}=Z_{e}\hat{n}\times [H_{2}-H_{1}]\\
        \hat{n}\times [H_{avg}]=Y_{m}\vec{M_{s}}=-Y_{m}\hat{n}\times [E_{2}-E_{1}]
    \end{aligned}
\end{equation}
where $Z_{e}$ is the electric surface impedance and 
$Y_{m}$ is the magnetic surface admittance equivalent to $1/Z_{m}$.
In fact, the electric and magnetic meta-atoms are each described 
by a surface impedance of $LC$ oscillating circuit 
containing inductance $L$ and capacitance $C$.
Mathematically, we can formulate the surface impedance of the
electric and magnetic meta-atom as
\begin{equation}
\begin{aligned}
\label{eq:impedance}
        Z_{e} =\left(\frac{2\pi f C_{e}-1}{(2\pi f)^{2}L_{e}C_{e}}\right)j,~
        Y_{m} =\left(\frac{1-(2\pi f)^2 L_{m}C_{m}}{2\pi f C_{m}}\right)j
\end{aligned}
\end{equation}
where $f$ indicates the resonant frequency.
Each meta-atom behaves as an $LC$ circuit when its resonant frequency 
$f$ matches the frequency of the incident wave. 
Mathematically, the resonant frequency is equivalent to
    $f=(2\pi \sqrt{LC})^{-1}$.

Given $Z_{e}$ and $Y_{m}$, we can formulate the transmission 
coefficient $T$ and reflection coefficient $\Gamma$ 
of a meta-atom pair:
\begin{equation}
\begin{aligned}
\label{eq:coefficient}
        T=\frac{4-Y_{m}\cdot Z_{e}}{(2+Y_{m}\cdot \eta)(2+Z_{e}/\eta))},~
        \Gamma=\frac{2(Z_{e}/\eta-Y_{m}\cdot \eta)}{(2+Y_{m}\cdot \eta)(2+Z_{e}/\eta)}
\end{aligned}
\end{equation}
where $\eta$ is the wave impedance in free space.
Hence, by changing the surface impedance $(Z_{e}, Y_{m})$, we precisely control the 
phase of the coefficients, creating an arbitrary phase shift on the incident 
wave \cite{epstein2016huygens}.

The excitation of the electric and magnetic surface currents, 
or, equivalently, the values of $Z_{e}$ and $Y_{m}$ is tuned by 
changing the capacitive or inductive loading of the meta-atoms
as shown in \Cref{eq:impedance}.
Hence, to make HMS reconfigurable, we load 
a voltage-controlled capacitor,
varactor diode,
on each meta-atom.
By applying voltage across each varactor, 
we can arbitrary change the surface impedance, or equivalently,
the phase of the transmission or reflective coefficient.
%


Since the electric and magnetic meta-atoms are superimposed on the surface, 
we dissect the equivalent circuit model 
for the electric and magnetic meta-atom individually.

\subsection{Magnetic Meta-atom}
\begin{figure}
\centering
\includegraphics[width=1\linewidth]{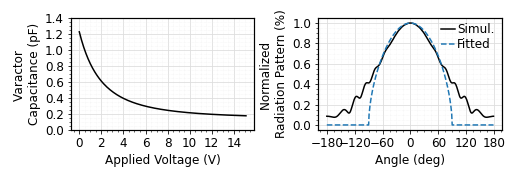}
\caption{\emph{Left:} $C_{var}$ as the voltage applied to varactor
changes, modeled with SPICE simulation; \emph{Right:} \shortname{}
element normalized beam pattern $F(\theta)$ simulated with HFSS and 
fitted function.}
\label{f:cvar_v_control}
\end{figure}

In this section, we provide the formulas for the magnetic meta-atom's capacitance and inductance
discussed in \cref{s:design:unit:params}.
First, we define the inductance of a circular metallic loop $L_{loop}$ as 
\begin{equation}
    L_{loop} = \mu_{0}R\left(log\left(\frac{8R_{m}}{t+w} - 
        \frac{1}{2}\right)\right),
\end{equation}
where $R$ is a mean radius, and $\mu_{0}$ is free-space permeability. 
Since there is a gap on the top of a metallic loop, 
the inductance of our magnetic meta-atom can be calculated as
\begin{equation}
    L_{m} = p_{m}L_{loop}=\left(1-\frac{g}{2\pi R}\right)L_{loop},
\end{equation}
where $g$ is a length of the gap. 
Now, we present the calculation of $C_{m}$.
First, the gap in the metallic loop creates a parallel-plate capacitance as follow:
\begin{equation}
    C_{gap} = \epsilon \frac{wt}{g}+\epsilon (t+w+g),
\end{equation}
where $w$ is the width of the loop, and $t$ is the thickness of the copper. 
Here, $\epsilon = \epsilon_{0}\epsilon_{eff}$ 
where $\epsilon_{0}$ is free-space permittivity,
and  $\epsilon_{eff}$ is effective permittivity, which can be calculated as
\begin{equation}
    \epsilon_{eff}=\frac{\epsilon_{r}+1}{2} + 
        \left(\frac{\epsilon_{r}-1}{2}\right)
        \left(\frac{1}{\sqrt{(1+12t/e)}}\right)
\end{equation}
where $\epsilon_{r}$ is the permittivity of the substrate. 
Second, there is a capacitance induced by the metallic ring itself: 
\begin{equation}
    C_{surf} = \frac{2\epsilon(t+w)}{\pi}ln\left(\frac{4R}{g}\right)
\end{equation}
Lastly, the varactor diode adds the capacitance as discussed in \cref{s:design:unit:params}. 
We have modeled our varactor, of Macom 
\href{https://www.macom.com/products/product-detail/MAVR-000120-14110P}{MAVR-000120-1411},
based on its \emph{Simulation Program with Integrated Circuit Emphasis} (SPICE) model 
and demonstrate our simulated $C_{var}$ values
in the left subfigure of \cref{f:cvar_v_control}.
Then, we formulate $C_{m}$ according to \cref{eq:magcap}.
Finally, model the circuit diagram as a series impedance where the 
series impedance itself corresponds to the surface impedance $Z_{m}=1/Y_{m}$.

\subsection{Electric Meta-atom}
Now, we provide the capacitance and inductance calculation for the electric meta-atom.
First, we formulate the inductance of a half-circle ring $L_{curve}$ as follow:
\begin{equation}
    L_{curve}=\left(p_{e}L_{circle}\right)/2 = 
        \frac{1}{2}\left(\left(1-\frac{g}{2\pi R_{m}}\right)L_{circle}\right).
\end{equation}
Based on \cite{esmail2019new}, we compute the the inductance of the strip as
\begin{equation}
\begin{array}{lr}
    L_{strip} = &\mu_{0}l/4\pi\left[2\sinh^{-1}(\frac{l}{w}) + 
    2(\frac{1}{w})\sinh^{-1}(\frac{w}{l}) - \right.\\
    &\left.\frac{2(w^{2}+l^{2})^{1.5}}{3lw^{2}} + 
    \frac{2}{3}(\frac{l}{w})^{2} + \frac{2}{3}(\frac{w}{l})\right]
\end{array}
\end{equation}
where $l$ is the length of strip, which is equivalent to $2R_{m}$, and $w$ 
is the width of the trace.
We then combine all inductance values into $L_{e}$ as
\begin{equation}
    L_{e} = (L_{curve}/2)+L_{strip}
\end{equation}
The formulas for the gap capacitance and surface capacitance for the electric meta-atom 
are the same as the magnetic meta-atom, and we define $C_{e}$ according to \cref{eq.elecap}.
Finally, the surface impedance of the electric meta-atom corresponds to a shunt impedance.


\section{Path Loss Model}
\label{s:path_loss_model}


This section presents a standard path loss model calculation
largely following the development in prior similar efforts
targeting lower frequencies \cite{9206044}, useful for our
purposes to establish the basic feasibility of our design
prior to hardware fabrication and full\hyp{}scale evaluation.

First let us assume that a transmitter directly communicates with a receiver. 
According to the Friis formula \cite{1697062}, 
the power intercepted by the receiving antenna with effective 
aperture $Ae_{R}$ and distance between transmitter and receiver $d$ is:
\begin{equation}
\label{eq:friis}
    P_{i}=S_{R}Ae_{R} = \left(\frac{P_{T}}{4\pi d^2} G_{T}\right)Ae_{R}
\end{equation}
where $S_{R}$ is the received power density, 
and $G_{T}$ is the peak gain of the transmitting antenna.
Since the effective aperture $Ae_{R}=\frac{\lambda^2}{4\pi}G_{R}$ 
where $G_{R}$ denotes the gain of the receiving antenna, we
rewrite \Cref{eq:friis} as
\begin{equation}
\label{eq:friis2}
    P_{i} = \left(\frac{P_{T}}{4\pi d^2} G_{T}\right)
    \left(\frac{\lambda^2}{4\pi}G_{R}\right) = 
    P_{T}G_{T}G_{R}\left(\frac{\lambda}{4\pi d}\right)^{2}.
\end{equation}
Now we consider a transmitter communicating
with the receiver via \shortname{}. 
Given \Cref{eq:friis2}, we formulate the power 
the $nm^{\mathrm{th}}$ meta\hyp{}atom captures
from the transmitter as
\begin{equation}
\begin{aligned}
\label{eq:pi}
    P_{nm}^{i} = P_{T}G_{T}G_{w}
    \left(\frac{\lambda}{4\pi d_{i,nm}}\right)^{2},
\end{aligned}
\end{equation}
where $G_{w}$ denotes the gain of the meta\hyp{}atom in the 
direction of the transmitter, and $d_{i,nm}$ is the distance 
between the transmitter and $nm^{\mathrm{th}}$ meta\hyp{}atom.
Similarly, we can calculate the power received by the receiving antenna 
from the $nm^{\mathrm{th}}$ meta\hyp{}atom as:
\begin{equation}
    \begin{aligned}
    \label{eq:ps}
    P_{R,nm} = P^{s}_{nm}G_{R}G_{w}
    \left(\frac{\lambda}{4\pi d_{s,nm}}\right)^{2},
    \end{aligned}
\end{equation}
where $G_{w}$ is the meta-atom gain scattered in the direction of 
the receiver, $d_{s,nm}$ is the distance between $nm^{\mathrm{th}}$ 
meta-atom to the receiver,
$P_{nm}^{s}$ is the power applied by each meta-atom, 
and $P_{nm}^{s}=P_{nm}^{i}\epsilon$. 
Here, $\epsilon$ accounts for the limited efficiency of meta-atom and 
insertion losses associated with components.
To simplify the formula, we assume $\epsilon=1$.
To calculate the power from the transmitter to the receiver, 
we then combine \Cref{eq:pi,eq:ps}:
\begin{equation}
\begin{aligned}
    P_{R,nm} = P_{T}G_{T}G_{R}\frac{G_{w}G_{w}}{d_{i,nm}^2 d_{s,nm}^{2}}
    \left(\frac{\lambda}{4\pi }\right)^{4}
    \label{eq:receivedpower}
\end{aligned}    
\end{equation}
Here, we emphasize that in the link budget, we must calculate the gain of \shortname{} twice,
one for receiving and another for transmitting. Hence, \cref{eq:receivedpower} has two $G_{w}$.
Since \shortname{} consists of a large array of meta-atoms, we can formulate the
total received power as a sum of the received powers from all  
meta-atoms as 
\begin{equation}
    P_{R} = \left|\sum_{n=1}^N\sum_{m=1}^M C_{nm} 
        \sqrt{P_{R,nm}}e^{j\phi_{nm}}\right|^2,
\end{equation}
where $C_{n,m}$ denotes the transmission or reflection coefficient of
the $nm^{\mathrm{th}}$ meta-atom,
and the phase $\phi_{nm}=2\pi(d_{i,nm}+d_{s,nm})/\lambda$. 
In a lens mode $C_{n,m} = T_{n,m}$, and in a mirror mode $C_{n,m} = \Gamma_{n,m}$.
We already defined $T_{n,m}$ and $\Gamma_{n,m}$ in eq.~\cref{eq:coefficient}.
Finally, we write the total received power as:
\begin{equation}
\begin{aligned}
    \label{eq:total_pr}
    P_{R} = P_{T}G_{T}G_{R}\left(\frac{\lambda}{4\pi}\right)^{4}
    \left| \sum_{n=1}^{N}\sum_{m=1}^{M}C_{nm} 
        \frac{\sqrt{G_{w}G_{w}}}{d_{i,nm}d_{s,nm}}e^{j\phi_{nm}}
    \right|^{2}.
\end{aligned}
\end{equation}
However, the meta-atom gain $G_{w}$ is unknown.
Thus, we re-define $G_{w}$ as a power radiation pattern from each 
meta-atom, which is equivalent to $GF(\theta_{nm})$. 
$G$ is a gain that depends on the physical area (\textit{i.e.} the effective aperture) of the meta-atom, 
and $F(\theta_{nm})$ is the normalized power radiation pattern.
Based on the effective aperture formula, 
$G=(4\pi/\lambda^2) Ae_{nm}=(4\pi/\lambda^2)(xy)$ where 
$x$ and $y$ are a vertical and horizontal meta-atom spacing, respectively.
Unlike traditional antennas with $x=y=\lambda/2$, 
our meta-atom has $x=\lambda/4.8$ and $y=\lambda/3.4$.
Moreover, $F(\theta_{nm})$ defines the variation of the 
power radiated or received by a meta-atom:
\begin{equation}
\begin{aligned}
    F(\theta) =
\begin{cases}
 cos^q(\theta)& \theta \in \left [ 0,\pi/2 \right ] \\
 0 & \theta \in \left [ \pi/2,\pi \right ]
\end{cases}
\end{aligned}
\label{eq:elementpattern}
\end{equation}
where $\theta$ are the angle from the meta-atom 
to a certain transmitting or receiving direction. 
In the right subfigure of \cref{f:cvar_v_control}, 
we present a simulated \shortname{} element beam pattern $F(\theta_{nm})$
as well as the curve fitted with \cref{eq:elementpattern}.
Based on our curve fit, $q=0.5611$.

\parahead{Far-field beamforming.}
In the far-field, we can approximate $d_{s,nm}=d_{s}$ and $d_{i,nm}=d_{i}$ 
since $d_{i}$ and $d_{s}$ are much greater than the distance between different meta-atoms.
However, we do not approximate $d_{s,nm}=d_{s}$ and $d_{i,nm}=d_{i}$ for the phase $\phi_{nm}$.
Then, we can simplify \Cref{eq:total_pr} as:
\begin{equation}
    \begin{aligned}
        \label{eq:total_pr2}
        P_{R} = P_{T}G_{T}G_{R}
        \left(\frac{Ae_{nm}}{4\pi d_{i}d_{s}}\right)^{2} 
        F(\theta_{i}) F(\theta_{s}) 
    \left| \sum_{n=1}^{N}\sum_{m=1}^{M}C_{nm} e^{j\phi_{nm}}
    \right|^{2}
    \end{aligned}
\end{equation}
This indicates that we can maximize the received power by configuring 
each meta-atom's $\angle C_{nm}$ to $-\phi_{nm}$.
Finally, the path loss of a correctly reconfigured \shortname{} as:
\begin{equation}
    \begin{aligned}
        \label{eq:total_pr3}
        L_{\shortname}^{-1} = 
        (\frac{xy}{4\pi d_{i}d_{s}})^{2} F(\theta_{i}) F(\theta_{s}) 
    \left| \sum_{n=1}^{N}\sum_{m=1}^{M}|C_{nm}|
    \right|^{2}
    \end{aligned}
\end{equation}
Since $0<|C_{nm}|<1$ for both transmissive and reflective mode, 
increasing the number of meta-atoms $N$ and/or $M$ reduces the path loss. 
Assuming $|C_{nm}|$ is close to $1$, the path loss of \shortname{} is 
proportional to $1/(NM)^2$.
While increasing the element spacing $x$ and $y$ seems to reduce the loss, 
it is not always true because $|C_{nm}|$ decreases 
when $x$ and $y$ increases due to increasing coupling between adjacent meta-atoms.

\section{Meta-atom controllability and sensitivity}
\label{s:vna}

\begin{figure}[t]
    \begin{subfigure}[b]{1\linewidth}
    \centering
    \includegraphics[width=.9\linewidth]{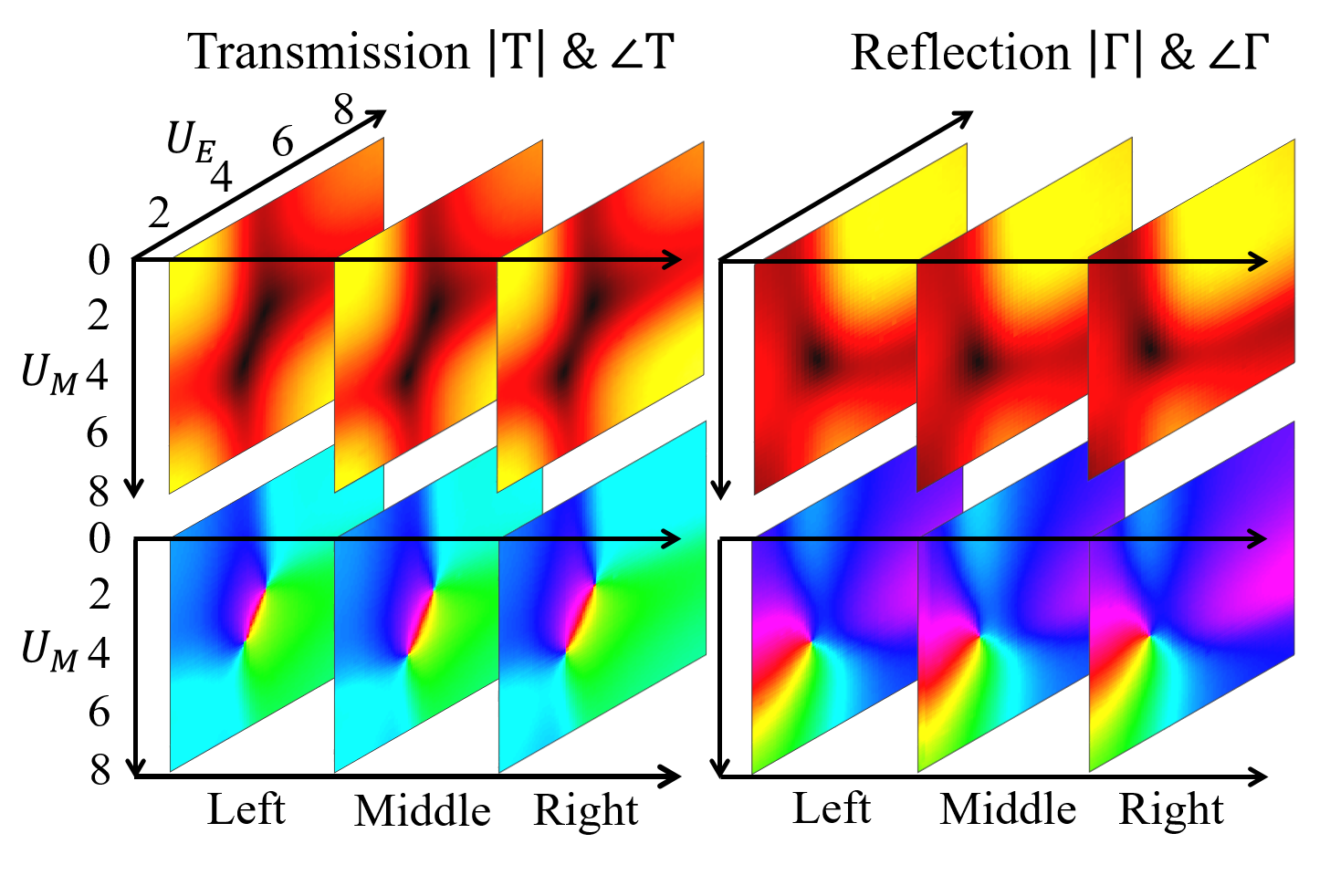}
    \caption{Sensitivity.}
    \label{f:eval:meta-atom}
    \end{subfigure}
    \begin{subfigure}[b]{1\linewidth}
    \centering
    \includegraphics[width=.9\linewidth]{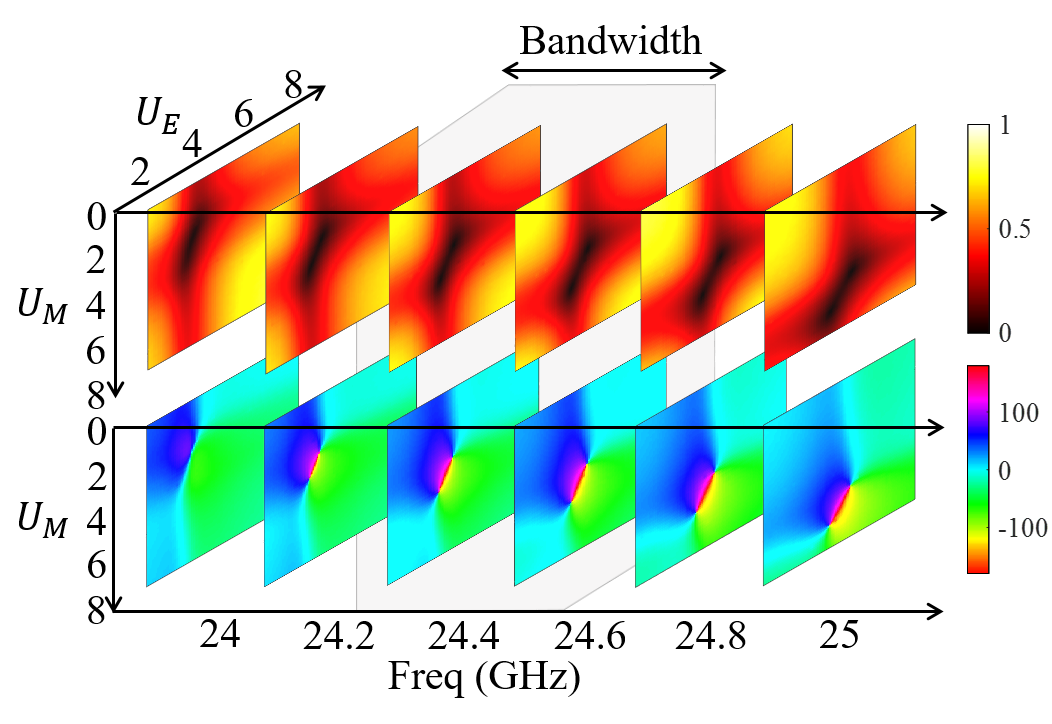}
    \caption{Bandwidth.}
    \label{f:huygen_bandwidth}
    \end{subfigure}
    \caption{Meta-atom microbenchmark}
\end{figure}

We present the Huygens pattern measured from the VNA in \cref{f:eval:meta-atom}. 
We measured the near-field Huygens pattern in three different areas of \shortname{} 
to evaluate \shortnames{} sensitivity against to fabrication variation.
For all three areas, we observe a $360$-degree phase variation with a high magnitude for both transmission and reflection. Moreover, the patterns do not vary across different area of the surface, signifying that the manufacturing tolerance did not greatly affect \shortnames{} near-field performance.
We also demonstrate the huygens pattern across \shortname{}'s operating bandwidth in \cref{f:huygen_bandwidth}.
Within the $200$ MHz bandwidth, the pattern is consistent.

\end{document}